\documentclass[aps, pra, twocolumn, final, a4paper, 10pt, numbers, showkeys, sort&compress]{revtex4-1}

\usepackage{amsmath}
\usepackage{amssymb}
\usepackage{graphicx}
\usepackage{color}
\usepackage{booktabs}
\usepackage{multirow}
\usepackage{dcolumn}
\usepackage{tabularx}
\usepackage{soul}
\usepackage{bbold}
\usepackage[colorlinks, urlcolor=black, citecolor=magenta, linkcolor=black]{hyperref}

\newcommand{\ket}[1]{|#1\rangle}
\newcommand{\bra}[1]{\langle#1|}
\newcommand{\g}{|g\rangle}
\newcommand{\e}{|e\rangle}

\newcommand{\gp}[1]{{\color{black}#1}}

\newcommand{\fs}[1]{{\color{black}#1}}

\newcommand{\fsB}[1]{{\color{black}#1}}

\begin{document}
\title{Fast and scalable quantum information processing with two-electron atoms\\in optical tweezer arrays}
\author{G. Pagano$^1$, F. Scazza$^{2,3}$, M. Foss-Feig$^4$}
\affiliation{$^1$Joint Quantum Institute, University of Maryland, College Park, MD 20742, U.S.A.}
\affiliation{$^2$ Istituto Nazionale di Ottica del Consiglio Nazionale delle Ricerche (INO-CNR), 50019 Sesto Fiorentino, Italy}
\affiliation{$^3$ LENS and Dipartimento di Fisica e Astronomia, Universit\`a di Firenze, 50019 Sesto Fiorentino, Italy}
\affiliation{$^4$ United States Army Research Laboratory, Adelphi, Maryland 20783, USA}

\begin{abstract}
Atomic systems, ranging from trapped ions to ultracold and Rydberg atoms, offer unprecedented control over both internal and external degrees of freedom at the single-particle level. They are considered among the foremost candidates for realizing quantum simulation and computation platforms that can outperform classical computers at specific tasks.
In this work, we describe a realistic experimental toolbox for quantum information processing with neutral alkaline-earth-like atoms in optical tweezer arrays. In particular, we propose a comprehensive and scalable architecture based on a programmable array of alkaline-earth-like atoms, exploiting their electronic clock states as a precise and robust auxiliary degree of freedom, and thus allowing for efficient all-optical one- and two-qubit operations between nuclear spin qubits. 
The proposed platform promises excellent performance thanks to high-fidelity register initialization, rapid spin-exchange gates and error detection in readout. 
As a benchmark and application example, we compute the expected fidelity of an increasing number of subsequent SWAP gates for optimal parameters, which can be used to distribute entanglement between remote atoms within the array.
\end{abstract}

\keywords{quantum information with ultracold atoms; spin-exchange quantum gates; optical tweezer arrays; two-electron atoms; clock state manipulation}

\maketitle
\section{Introduction}

During the last two decades, neutral atoms have been explored  as an efficient and scalable quantum computing platform \cite{Jaksch1999, Brennen1999, Jaksch2000, Briegel2000, Calarco2000, Lukin2001, Calarco2004, Hayes2007}. In these systems, internal atomic states with very long coherence times are typically available, allowing for the robust storage of quantum information. By applying suitably tailored radiation, single qubits can be manipulated with fidelities above 99$\%$, as has been demonstrated in numerous experiments \cite{Saffman2010,Xia2015,Wang2016}. While large numbers of qubits can be initialized, read out, and coherently manipulated at the single-particle level,  the implementation of scalable,  high-fidelity two-qubit entangling gates, often realized through the Rydberg blockade effect \cite{Maller2015, Levine2018}, remains an outstanding challenge.

In this context, collisional two-atom gates present key advantages, as they are intrinsically local, thereby reducing complexity and enhancing scalability \cite{DiVincenzo2000, Mandel2003, Ratcliffe2018}. Specifically, entangling gates for trapped atoms have been proposed based on spin exchange, arising from spin-dependent collisional dynamics \cite{Hayes2007, Weitenberg2011a}. However, despite impressive proof-of-principle implementations \cite{Anderlini2007, Kaufman2014, Kaufman2015}, 
no complete spin-exchange-based universal computing architecture has been implemented to date.
The aim of this proposal is to connect well-established ideas in quantum information processing with neutral atoms with state-of-the-art experimental capabilities.
In this work, we propose and analyze a realistic architecture based on spin-exchange entangling gates among neutral alkaline-earth-like (AEL) atoms trapped in programmable optical tweezers, allowing gate fidelities above 99\% and  favorable scaling to large arrays of qubits. 

Research with ultracold gases of AEL atoms has expanded rapidly in recent years, motivated by their unique and versatile internal structure that offers exciting possibilities for quantum simulation \cite{Gorshkov2010, Gerbier2010, Cazalilla2014} and quantum metrology \cite{Ludlow2015}. Owing to the simultaneous presence of nuclear and electronic degrees of freedom with exceptional coherence properties, fermionic two-electron atoms provide a powerful toolbox for quantum information with ultracold atomic gases \cite{Daley2008, Gorshkov2009, Shibata2009, Daley2011b}, that makes them ideal candidates for realizing scalable quantum computing architectures. 
In particular, $^{173}$Yb and $^{87}$Sr feature the combined presence of a long-lived $^3P_0$ excited state and the associated ultranarrow $^1S_0\rightarrow {^3}P_0$ clock transition, with a recently discovered large and elastic spin-exchange coupling between the two clock states \cite{ Scazza2014, Zhang2014a, Cappellini2014a, Goban2018}, a key resource for the implementation of efficient spin-exchange-based quantum gates. 
Moreover, the nuclear spin degree of freedom of AEL atoms is particularly suitable for encoding quantum information, owing to its weak sensitivity to external magnetic and electric fields, and to SU($N$) symmetry of collisions for both clock states ($J=0$) \cite{Pagano2014, Scazza2014, Zhang2014a}, which increases the flexibility of quantum information schemes \cite{Gorshkov2009}. Such exceptional features have already been exploited to realize the most precise atomic clocks to date \cite{Hinkley2013, Bloom2014, Campbell2017}, and for quantum simulation of many-body physics in novel regimes \cite{Mancini2015, Hofrichter2016, Livi2017, Riegger2018}.

In this proposal, we show that programmable arrays of optical tweezers, which have recently stimulated intense research efforts \cite{Kaufman2014, Kaufman2015, Endres2016, Labuhn2016, Barredo2016, Bernien2017,Levine2018, Liu2018, Ni2018}, allow for optimally harnessing the spin-exchange coupling between the clock states of $^{173}$Yb or $^{87}$Sr to implement a fully controllable quantum register with practical individual-qubit addressing and detection, great robustness and fast repetition rates. 
The paper is organized as follows. In Section \ref{secarch}, the general architecture based on AEL atoms in programmable optical tweezers will be outlined. In Section \ref{sec_gates}, we explain how the electronic clock state can be used as a convenient auxiliary degree of freedom to perform efficient one-qubit gates and to realize fast and reliable two-qubit entangling operations via spin-exchange interactions. In Section \ref{sec_reg_in}, we discuss the experimental techniques to cool down, initialize, and read out the atomic array. 

\section{Architecture\label{secarch}}
\begin{figure}[t!]
\centering
\includegraphics[width=1\columnwidth]{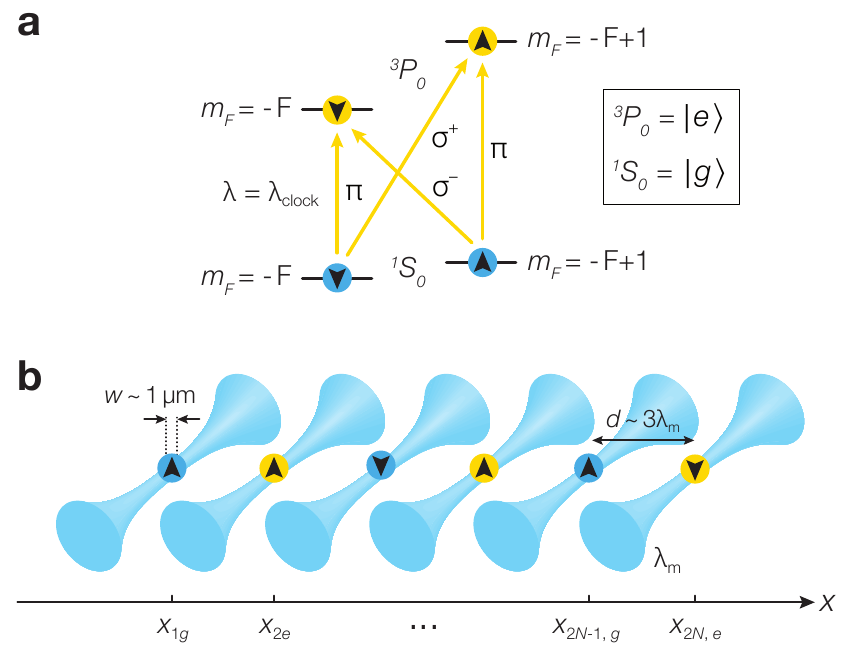}
\caption{{\bf General architecture for quantum information with neutral AEL atoms in optical tweezer arrays.} ({\bf a}) Qubit encoding in the nuclear spin states and relevant clock transitions. The auxiliary optical clock qubit is exploited to perform one-qubit rotations, two-qubit spin-exchange gates and shelved detection. ({\bf b}) Staggered $\ket{g}$-$\ket{e}$ configuration in a magic-wavelength one-dimensional tweezer array.}
\label{fig_1}
\end{figure}
The general concept of our scheme is illustrated in Fig.~\ref{fig_1}. The starting point of this proposal is a one-dimensional array of optical tweezers loaded with single $^{173}$Yb or $^{87}$Sr atoms \cite{Kaufman2014, Kaufman2015, Endres2016, Bernien2017}. These atoms endow both long-lived electronic states and magnetically insensitive nuclear spin states \cite{Daley2011b}. The latter can be conveniently used as a memory qubit (see Fig.~\ref{fig_1}a), featuring very long coherence times owing to their natural resilience to environmental magnetic perturbations, a feature shared with other promising quantum memories \cite{Olmschenk2007}. On the other hand, the electronic degree of freedom serves as an ancillary qubit to perform both one-qubit coherent manipulations and measurements, as well as state-selective atom transport \cite{Daley2008, Daley2011a,Daley2011b}. In the case of $^{173}$Yb and $^{87}$Sr, the upper $\ket{^3P_0}\equiv\ket{e}$ state is connected to the ground $\ket{^1S_0}\equiv\ket{g}$  state by ultranarrow \emph{clock} transitions at 578\,nm and 698\,nm, respectively. The lifetime of the $\ket{e}$ state, $\tau > 20$\,s, allows for several coherent single-photon manipulations of the optical qubit state within reasonable experimental timescales. Our protocol exploits deep magic-wavelength tweezers at $\lambda_{m}\simeq 465$\,nm for $^{173}$Yb and $\lambda_{m}\simeq 497$\,nm  for $^{87}$Sr, where the $\ket{g}$ and $\ket{e}$ states possess equal AC polarizabilities (see Table~\ref{table_1}-\ref{table_2}). In such magic traps, the manipulation of the optical qubit becomes independent of the trapping potential, and atoms in different clock states experience equal potentials when merged into the same single-tweezer trap. 

For an efficient implementation of two-qubit operations based on the spin exchange between $\g$ and $\e$ states (see Section~\ref{sec_gates}), the register needs to be initialized in a staggered electronic-state configuration (see Fig.~\ref{fig_1}b): odd (even) traps are occupied by $\g$-atoms ($\e$-atoms). 
This ensures that only a single $\g$- and $\e$-state atom participate in a spin-exchange gate between neighbouring atoms, completely avoiding inelastic lossy collisions between $\ket{e}$ atoms \cite{Scazza2014, Zhang2014a}, which is a fundamental limitation of other schemes \cite{Gorshkov2009, Daley2008}.
We will discuss the initialization of the register in this alternated configuration in Section~\ref{sec_reg_in}.

Computational qubits are encoded in the atomic nuclear spin degree of freedom $I$, with $F=I=5/2$ for $^{173}$Yb and $F=I=9/2$ for $^{87}$Sr, for both the $\ket{g}$ and $\ket{e}$ states. For simplicity, we will limit the occupation of the nuclear spin manifold containing $2F+1$ states to only two adjacent spin states, which are taken as $m_F=-F,-F+1$ in the following without loss of generality (see Fig.~\ref{fig_1}a). The computational basis is therefore given by:
\begin{align*}
\label{eq:comp_basis}
\ket{0} &= \ket{\!\downarrow} \equiv \ket{m_F=-F}\\
\ket{1} &= \ket{\!\uparrow} \equiv \ket{m_F=-F+1}.
\end{align*}

\begin{figure}[b!]
\centering
\includegraphics[width=\columnwidth]{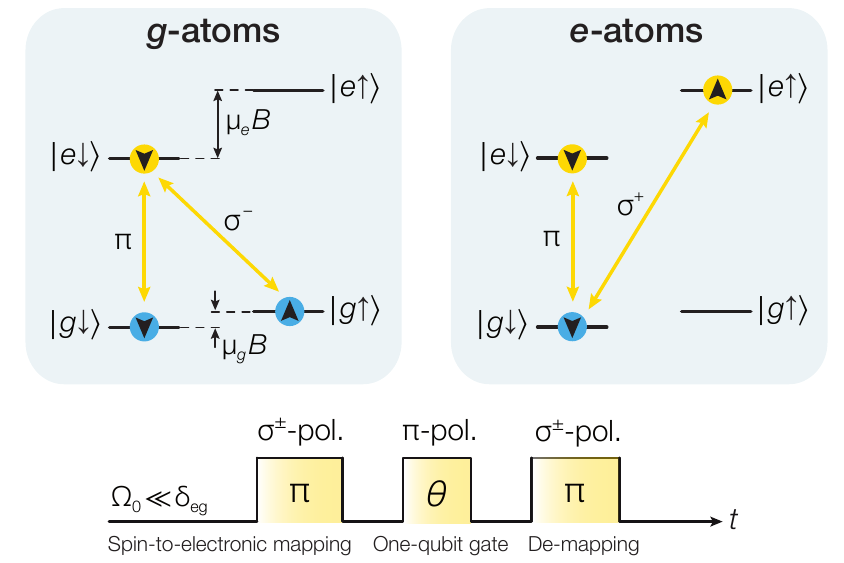}
\caption{{\bf Clock pulse sequence for arbitrary one-qubit operations}. Using two spin-selective global mapping $\pi$-pulses with $\sigma^\pm$ enclosing the one qubit $\pi$-polarized pulse, arbitrary one-qubit rotations can be performed.}
\label{fig_2}
\end{figure}
\begin{figure*}[t]
\centering
\includegraphics[width=\textwidth]{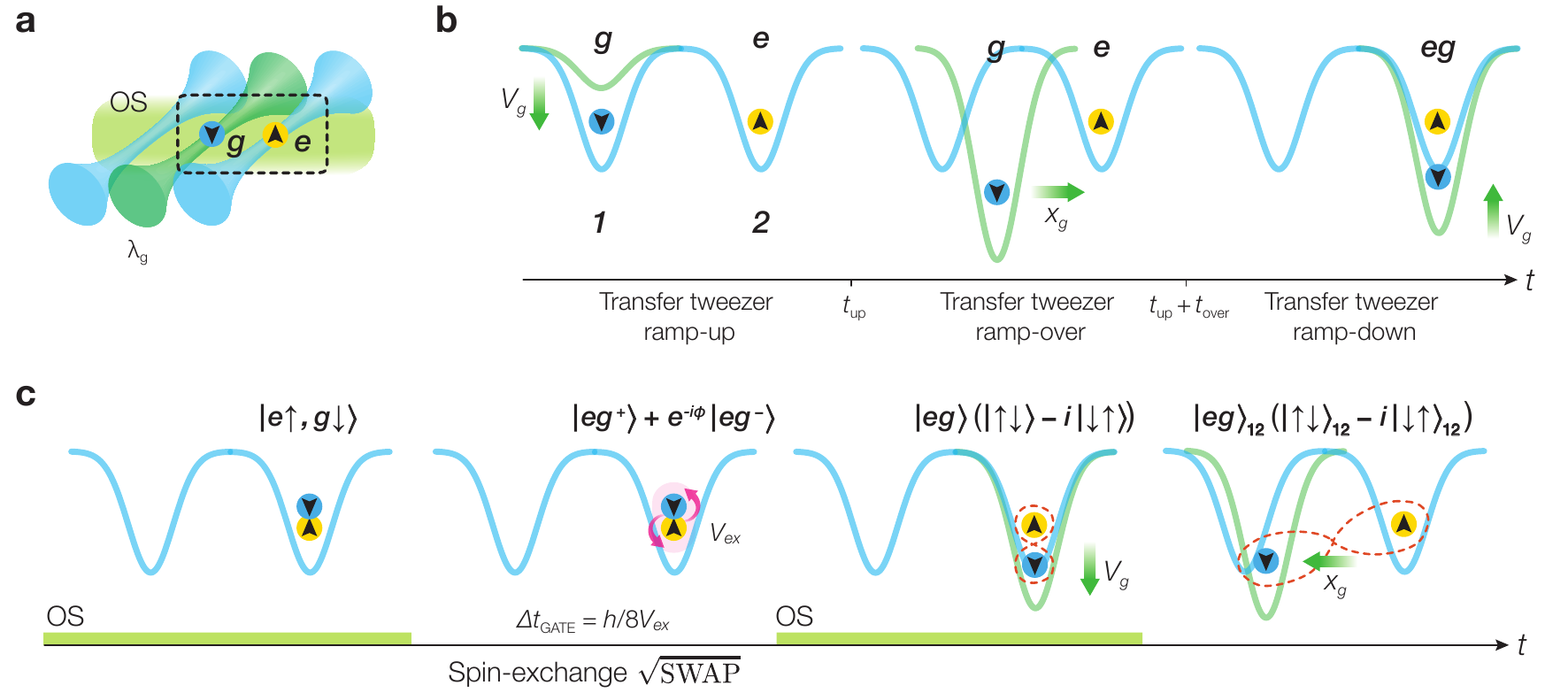}
\caption{{\bf Spin-exchange two-qubit gate protocol}. \textbf{(a)} A transport tweezer at the tune-out wavelength $\lambda_g$ is used to adiabatically transfer a $\g$-state atom into the adjacent magic-wavelength tweezer, occupied by a $\e$-state atom. 
An OS beam is used to provide a strong spin-dependent light shift, so as to energetically inhibit the spin-exchange process between the two atoms. \textbf{(b)} The transport tweezer is first ramped up to a depth $V_g$ in a time $t_\textrm{up}$, then its position is shifted to the position of the neighboring trap in a time $t_\textrm{over}$, and finally it is ramped down to 0 in a time $t_\textrm{down}$. \textbf{(c)} A $\sqrt{\textrm{SWAP}}$ gate between states $\ket{e\uparrow}$ and $\ket{g\downarrow}$ is performed by turning off and on the OS beam for a time $\Delta t_\textrm{gate} = h/8V_{\rm ex}$, while the two atoms are trapped in the same magic tweezer. During this time, half a spin-exchange oscillation occurs, entangling the two nuclear spin qubits.
}
\label{fig_3}
\end{figure*}
%
In the staggered array, fast two-qubit entangling exchange gates are implemented by using a moving \emph{transport} tweezer beam at the tune-out wavelength $\lambda_g \simeq 570$\,nm for $^{173}$Yb or $\lambda_g \simeq 633$\,nm for $^{87}$Sr to selectively merge/de-merge $\g$-atoms with $\e$-atoms occupying a neighboring potential well. The exchange gate can be timed precisely by exploiting a strong spin-dependent light shift from a circularly polarized beam, closely detuned from the $^1S_0\rightarrow{^3P}_1$ transition (see Section~\ref{sec_2qubit} for details).

\section{Gate operations}\label{sec_gates}

\subsection{One-qubit gate}

One-qubit gates are performed through a composite clock-pulse sequence (see Fig.~\ref{fig_2}). The following scheme is used for $\ket{g}$ atoms (odd traps):
$(1)$ the $\ket{g\!\!\uparrow}$ population is mapped on a $\ket{e\!\!\downarrow}$ with a $\sigma^-$-polarized clock $\pi$-pulse, mapping the nuclear qubit on the optical qubit ($\ket{g}=\ket{0},\, \ket{e}=\ket{1}$);
$(2)$ we perform the gate on the optical qubit by driving the clock transition at a Rabi frequency $\Omega_0$ using $\pi$-polarized light; $(3)$ we re-map the optical qubit on the nuclear qubit $\ket{e\!\downarrow}~\rightarrow~\ket{g\!\uparrow}$ with a second $\sigma^-$-polarized clock $\pi$-pulse.
For $\ket{e}$ atoms (even traps), the same procedure applies, but the nuclear-optical qubit mapping is performed with a $\sigma^+$ clock pulse.
\begin{figure*}[t!]
\centering
\includegraphics[width=\textwidth]{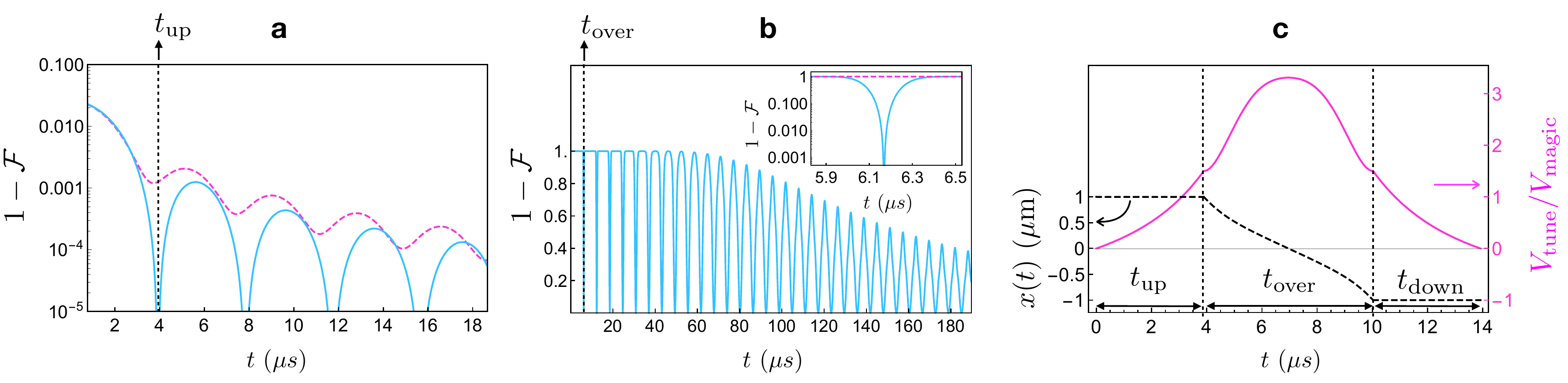}
\caption{{\bf Ground state fidelity during transport for two-qubit gates.} (\textbf{a}) Motional ground state infidelity after a ramp up of the transport tweezer in time $t$ to a final depth of $V_g(t)=1.5\,V_{\rm magic}$. Purple dashed line (blue solid line) is computed using the naive (locally adiabatic) protocol. (\textbf{b}) Motional ground state infidelity after the transport tweezer is swept over to the the adjacent tweezer, assumed to be separated by 2 $\mu$m (locally adiabatic only).  In the inset, one can see that, even at very short times, high-fidelity diabatic transport is possible. 
At such short times the naive protocol (inset, purple dashed line) yields essentially zero final ground-state population. (\textbf{c}) Intensity and position of the transport tweezer required to achieve a locally adiabatic ramp.  The profile shown gives rise to a final infidelity of order $10^{-3}$ in a time $t\sim 14\,\mu$s.  All plots are for $^{173}$Yb, with a magic wavelength of $465.4$ nm and $\omega_{\rm tw} \simeq 2\pi \times 103$ kHz.}
\label{fig_4}
\end{figure*}
The speed of one-qubit operations is determined by the clock Rabi frequency $\Omega_0$. The frequency selectivity between $\pi$ or $\sigma^\pm$ clock transitions associated to different $m_F$ states is only limited by the differential energy shift $h \delta_{eg}$ between the $\g$ and the $\e$-state manifolds. In order to shift the different $m_F$ states within the $\g$-state manifold from one another, a spin-dependent light shift can be applied by shining a circularly-polarized beam closely detuned from the $^1S_0\rightarrow{^3P}_1$ transition. An energy separation between the $\ket{\!\!\downarrow}$ and $\ket{\!\!\uparrow}$ of the order of $h \times 1\,$MHz is easily achievable (see Appendix~\ref{app_B}), and allows in principle for $\mu$s-scale one-qubit operations. In addition, a quantization magnetic field $B$ can be used to induce a differential Zeeman shift $\delta_{eg} = m_F B \times 112(1)\,\text{Hz/G} - B^2 \times 0.062(1)$\,Hz/G$^2$ for$^{173}$Yb \cite{Porsev2004, ScazzaPhD}, and $\delta_{eg} =  m_F B \times 108.4(4)\,\text{Hz/G} - B^2 \times 0.233(5)$\,Hz/G$^2$ for $^{87}$Sr \cite{Boyd2007}.


\subsection{Two-qubit exchange gate}\label{sec_2qubit}
The two-qubit gate is based on coherent spin-exchange interactions between $\g$ and $\e$ atoms \fsB{occupying the motional ground state of the same tweezer}. As shown in \cite{Pagano2015a, Cappellini2014a, Scazza2014,Zhang2014a,Goban2018}, atoms in different electronic states $\ket{g}$ and $\ket{e}$ feature a strong spin-exchange interaction $V_{\rm ex}$, whose strength is set by the difference between the triplet and singlet scattering lengths $a_{eg}^\pm$ \cite{Gorshkov2010}. Two atoms can be entangled using a $\sqrt{\rm SWAP}$ gate, realized by adiabatically merging the atoms and letting them undergo spin-exchange dynamics for a time $\Delta t_{\rm gate}=h/(8V_{\rm ex})$. 

\subsubsection{Perturbative interactions}

In this section we will assume that the interactions can be treated perturbatively, which is justified as long as both $a_{eg}^{\pm}$ are small compared to the width of the non-interacting ground-state wavefunctions for both $\g$ and $\e$ atoms.  This approximation is appropriate for $^{87}$Sr, but may break down for $^{173}$Yb; effects of the breakdown of perturbation theory on the exchange gate will be discussed in Section \ref{sec:strong_interactions}.

The process of adiabatically transferring two atoms that are initially in different tweezers into the same tweezer is rather slow if the potentials are state-independent, ultimately being limited by energy scales much smaller than the trap frequency (e.g., the tunneling energy between two nearby tweezers \cite{Kaufman2015}).  Here, in addition to the background magic-wavelength potential, we employ an additional far-off resonant, state-dependent transport tweezer that only traps atoms in the $\ket{g}$ state, with a time-dependent depth $V_g(t)$ and central position $x(t)$ (measured relative to the center of the magic-wavelength tweezer trapping the $g$-atom). For two neighboring atoms, one in $\ket{g}$ and one in $\ket{e}$, a possible approach to merging their wavefunctions for the application of a two-qubit gate is as follows (see Fig.\ \ref{fig_3}b): (1) For a time $t_{\rm ramp}$ the confinement of the $\ket{g}$ atom is increased by adiabatically ramping up the transport tweezer depth $V_g(t)$. (2) The center of the confining potential for the $\ket{g}$ atom is then shifted over (in a time $t_{\rm over}$) until it coincides with the center of the tweezer containing the $\ket{e}$ atom. (3) The additional potential $V_g(t)$ is then adiabatically turned off in a time $t_{\rm down}$, leaving both atoms trapped in the motional ground state of the same magic-wavelength tweezer. This protocol does not require any movement of the magic-wavelength tweezers; in this way, it also avoids detrimental amplitude modulations of the trapping potentials, which would arise when initially separated magic-wavelength tweezers (closely detuned from one another) spatially overlap.

To prevent the premature initiation of spin-exchange during the end of step (2) and step (3) (during which the atomic wavefunctions are at least partially overlapped), a spin-dependent light shift can be applied via an optical switch (OS) beam (see Fig. \ref{fig_3}c). Note that the OS beam can be turned on and off extremely quickly compared to the (still fast) typical exchange timescale $t_{\rm ex}=h/V_{\rm ex}$, and with precise timing.  This feature is enabled by the use of electronic (rather than vibrational \cite{Anderlini2007,Kaufman2015}) states in the exchange process. In addition to enabling precise control over the onset of exchange oscillations, the application of an OS beam until the transport tweezer is fully ramped down ensures that pointing fluctuations in the tweezer beams generate only common-mode motional noise, and will not affect the spin-exchange rate.

The transport dynamics is straightforward to simulate because the atoms do not interact.  The details of the ramp can be optimized if desired, but here we simply give two illustrative examples.  In the first, which we call ``naive'', the transport-tweezer intensity is ramped up and down linearly in steps (1) and (3), respectively, and its center is moved at a constant velocity while maintaining a fixed intensity in step (2).  In the second, which we call ``locally adiabatic'', both $V_g(t)$ and $x(t)$ are manipulated during all three steps in such a way that during steps (1) and (3) the central position of the $\ket{g}$ atom potential is held fixed while the confining potential is increased, and in step (2) the $\ket{g}$ atom confinement is held fixed while the central position is moved.  Moreover, in each step the changes to the approximately harmonic confining potential are made locally adiabatic in the sense described in Ref.\ \cite{Weitenberg2011a}.

In Fig.\ \ref{fig_4} we show the ground-state infidelity $1-\mathcal{F}$ achieved by initiating the $g$-atom in its motional ground state, and then carrying out either step (1) [Fig.\ \ref{fig_4}(a)] or step (2) [Fig.\ \ref{fig_4}(b)], using both the naive (purple dashed line) and locally adiabatic (blue solid line) protocols. [Step (3) is not shown, as it is simply the time reversed process to step (1).] Here, $\mathcal{F}\equiv|\bra{\Psi_0(t)}\Psi(t)\rangle|^2$, where $\ket{\Psi_0(t)}$ is the instantaneous motional ground state at the end of the ramp and $\ket{\Psi(t)}$ is the wavefunction at the end of the ramp computed exactly within the harmonic approximation (see Appendix \ref{app_A} for details of the calculation).  While the envelopes are similar for both approaches, the locally-adiabatic approach shows much more pronounced near-zeros of the infidelity.  With this approach, we find that the entire transport process can in principle be carried out in a time $t\sim 3 \cdot 2\pi/\omega_{\rm tw}$ with an infidelity on the order of $10^{-3}$. However transporting an atom in such a short time while keeping the error at the $10^{-3}$ level requires a timing precision below 10\,ns, as one needs to exploit the first narrow minima of the transport infidelity [see Fig.\ (\ref{fig_4}(b)]. In addition, transport on this timescale will involve spatial excursions of the atom relative to the beam center for which a harmonic approximation of the tweezer potential is not strictly valid (see Appendix \ref{app_A} for more detail). In order to counteract possible residual heating into excited motional states of the transport tweezer arising during the transfer procedure, nuclear spin coherence-preserving sideband cooling of $\g$-state atoms could be performed in addition~\cite{Reichenbach2007}.

\begin{figure}[t!]
\includegraphics[width=0.75\columnwidth]{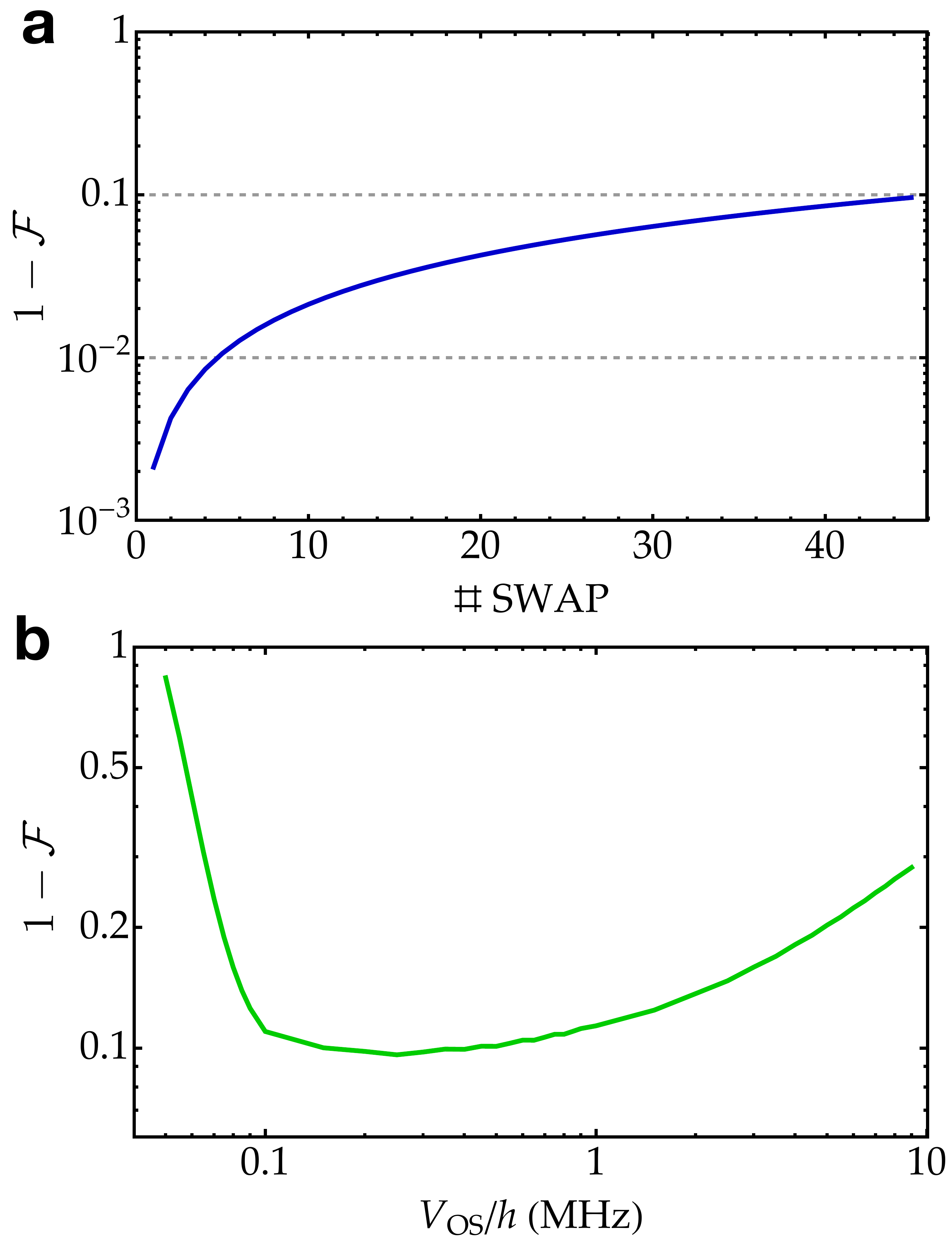}	
\caption{
{\bf Infidelity of repeated two-qubit $\rm SWAP$ exchange gate for $^{87}$Sr}. (\textbf{a}) Infidelity as a function of the number of $\rm SWAP$ gates, interleaved by intervals of transport $2 t_{\rm over} = 12\,\mu\rm s$ plus evolution under light shifts of duration $2 t_{\rm down}=6 \,\mu{\rm s}$. The parameter values are: $V_{\rm OS}/h =300~\mbox{kHz}$, $\omega_{\rm tw}/2\pi= 103~\mbox{kHz}$ \fs{and $V_{\rm ex}/h=6\,{\rm kHz}$}. The Raman, Rayleigh and motional heating rate are $\Gamma_{\uparrow\downarrow} = 1~\mbox{s}^{-1}$, $\Gamma_{el} = 0.1~\mbox{s}^{-1}$, $\eta^2 \Gamma_{\uparrow} = 6.3~\mbox{s}^{-1}$, respectively, with $\eta$ being the Lamb-Dicke parameter (see Appendix \ref{app_master} for details). We assume the infidelity of a single transport operation to equal $10^{-3}$ (see Fig.~\ref{fig_4}). A single SWAP operation with OS on during $2 t_{\rm down}$ introduces an error of $10^{-4}$. The horizontal dashed lines indicate the infidelity thresholds of 1\% and 10\%. (\textbf{b}) Calculated infidelity for 40 consecutive $\rm SWAP$ gates, as a function of the OS beam differential light shift $V_{\rm OS}$. For the parameters above, the optimal $V_{\rm OS}/h$ is about 250\,kHz.
}
\label{fig_5}
\end{figure}
After the transport process, both $g$ and $e$ atoms (approximately) occupy the non-interacting ground state of the same magic-wavelength tweezer. Because the motional states are identical, any two-atom state can be labeled unambiguously by its internal state, which must have the proper Fermi-antisymmetrization.  The most general wavefunction can therefore be written as
\begin{equation*}
\label{eq_basis}
\ket{\psi}=c_1\ket{g\!\downarrow,e\!\downarrow}+c_2\ket{g\!\uparrow,e\!\downarrow}+c_3\ket{g\!\downarrow,e\!\uparrow} + c_4 \ket{g\!\uparrow,e\!\uparrow}.
\end{equation*}
Here the $c_i$ denote the complex amplitudes of the four two-qubit states $\ket{g\sigma,e\sigma'}$ ($\sigma,\sigma'\!\!=\,\uparrow,\downarrow$), which in first quantization are the explicitly antisymmetrized states $\ket{g\sigma,e\sigma'}=(\ket{g\sigma}_{1}\ket{e\sigma'}_{2}-\ket{e\sigma'}_{1}\ket{g\sigma}_{2})/\sqrt{2}$. The dynamics of the above two-qubit state $\ket{\psi}$ is governed by the Hamiltonian
\begin{equation}
\label{eq_H_eg}
\hat{H}_{eg}=
\begin{pmatrix}
U_{eg}^{-}  + V_{\rm OS} (t) & 0 					& 0 				& 0\\
  0   &V_{\rm d } & V_{\rm ex }  			& 0\\
  0   &V_{\rm ex } & V_{\rm d} + V_{\rm OS} (t)& 0\\
  0   &0 & 0& U_{eg}^{-} \\
\end{pmatrix}.
\end{equation}
Here, $U_{eg}^{\pm}$ is the interaction energy for two atoms in the triplet ($+$) or singlet ($-$) electronic state. $V_{\rm OS}(t) = V_{\downarrow}-V_{\uparrow}$ is the differential spin-dependent light shift, while $V_{\rm d}=(U_{eg}^+ + U_{eg}^-)/2$ and $V_{\rm ex}=(U_{eg}^+ - U_{eg}^-)/2$ are the direct and exchange interaction energies, respectively.  Once the spin-dependent light shift is switched to $V_{\rm OS} = 0$, the two-qubit evolution associated with the Hamiltonian \eqref{eq_H_eg} after a time $\Delta t_{\rm gate}$ can be decomposed into a standard $\sqrt{\textrm{SWAP}}$ gate and a phase gate, and it can be used to generate maximally entangled two-qubit states between adjacent atoms. Entangled states between remote atoms can readily be realized through entanglement swapping, performing sequential SWAP gates between pairs of neighbouring atoms.



In order to estimate the gate fidelity the main sources of errors need to be considered, namely: $(i)$ imperfect suppression of the spin-exchange oscillations by the optical switch $V_{\rm OS}$, $(ii)$ motional decoherence due to elastic photon scattering events, and $(iii)$ spin decoherence caused by spontaneous scattering processes from the OS beam.

\

\begin{quote}
$(i)$ During the final transport stages (once the atoms begin to overlap) the spin population will undergo off-resonant exchange oscillations with amplitude $\sim 4 V_{\rm ex}^2/V_{\rm OS}^2$, inducing unitary errors in the spin exchange. Pointing fluctuations in the transport beam during this time will cause fluctuations in $V_{\rm ex}$, leading to further \emph{irreversible} degradation of the gate fidelity. Note that the OS-induced relative phase accumulation (during the ramp down) between different states in the computational basis must be known modulo $2\pi$.
\

$(ii)$ Motional heating will arise due to spontaneous photon scattering from the near-resonant OS beam, and to a lesser extent from the other far-off-resonant lasers. Assuming ground-state cooled atoms (see Section~\ref{sec_reg_in}), the photon scattering rate of the ${\rm OS}$ beam on the blue sideband is $\eta^2 \Gamma_{\sigma}$, where $\sigma=\uparrow,\downarrow$ and $\eta=\sqrt{\omega_{\rm R}/\omega_{\rm tw}}$ is the Lamb-Dicke parameter, with $\omega_{R}$ the photon recoil frequency and $\omega_{\rm tw}$ the harmonic oscillator frequency of the magic-wavelength tweezer potential. 
A single motional excitation significantly modifies the exchange interaction energy and would therefore spoil the gate. However this error source can in principle be suppressed by confining the atoms in deeper tweezers, \gp{at the expenses of a higher off-resonant scattering rate.}
\

$(iii)$ Scattering from the near resonant OS beam can also lead directly to spin state decoherence. However the only photon scattering events inducing decoherence are those that carry away information about the two qubit state \cite{Ozeri2007}. These can be of two types (see Fig. \ref{fig_6} in Appendix \ref{app_B}): Raman scattering events with rate $\Gamma_{\uparrow\downarrow}$, where a spin flip is involved, and elastic Rayleigh events, which cause pure dephasing at a rate $\Gamma_{\rm el}$ proportional to the square of the difference between the two spin state's elastic scattering amplitudes \cite{Uys2010}. 

\end{quote}

All processes can be accounted for using a master equation
\begin{equation}
\partial_t \rho=-\frac{i}{\hbar} [H,\rho] + \mathcal{L}(\rho),
\label{eq_master_equation}
\end{equation}
where $\rho$ is the density matrix and $\mathcal{L}$ is a Lindbladian superoperator (see Appendix \ref{app_C} for details). \fsB{The detrimental effect of motional excitations resulting from imperfect ground-state cooling is separately treated in detail in} Appendix~\ref{app_D}.

A trade off exists between these different errors for a given trapping frequency $\omega_{\rm tw}$: Those of type $(i)$ can be suppressed by increasing the OS laser power, at the expense of a larger motional heating rate $(ii)$ and faster direct spin decoherence $(iii)$.  By simulating the spin-exchange gate dynamics with Eq.~(\ref{eq_master_equation}), it is possible to determine the optimal OS beam parameters. The infidelity of a single $\rm SWAP$ operation due to these errors can readily be suppressed below $10^{-3}$, and the gate degradation due to decoherence over repeated gates is largely dominated by transport infidelities. As shown in Fig.~\ref{fig_5}, assuming $10^{-3}$ transport infidelity, the total gate error is predicted to reach about $10\%$ after 45 gates.

So far our analysis has been limited only to coherent exchange dynamics between $\ket{e\!\downarrow,g\!\uparrow}$ and  $\ket{e\!\uparrow,g\!\downarrow}$. 
In this case, the phase accrued because of direct interaction $V_{\rm d}$ during the gate protocol, $\phi_{\rm d}(t)\sim \int_{0}^t V_{\rm d}(t') dt'$, is an irrelevant global phase. However, when two atoms in the same spin state are merged into the same tweezer, they experience an interaction energy $U_{\rm eg}^{-}(t)$ (with time dependence during the transport procedure) purely from the electronic-state singlet channel, and acquire a collisional phase shift $\phi_{\rm s}(t)\sim \int_{0}^t U_{\rm eg}^{-}(t') dt'$ [see Eq.\ (\ref{eq_H_eg})]. Therefore, considering the full computational basis, the differential phase $\phi_{\rm s}-\phi_{\rm d}$ is relevant and is sensitive to the details of the transport procedure, including intensity and beam pointing fluctuations.  

In this context, fractional shot-to-shot intensity fluctuations $\delta I/I = \epsilon$ will yield a gate infidelity of order $\epsilon$. A relative intensity noise $\epsilon \lesssim 10^{-4}$ is typically achieved for actively stabilized optical traps, and the associated infidelity should therefore contribute less than other error sources discussed above.  Additionally, beam pointing fluctuations $\delta{x_{\rm c}}$ will induce relative phase fluctuations $\delta\phi_{\rm s,d}/\phi_{\rm s,d}\lesssim \delta{x_{\rm c}}/a_{\rm tw}$, with $x_{\rm c}$ being the tweezer center and $a_{\rm tw}$ being the tweezer harmonic oscillator length. 
For a vibrational gap of $\omega_{\rm tw}/2\pi \simeq 100\,\rm kHz$, a transport tweezer pointing stability of $\sim 3$\,nm is required to stabilize the differential phase $\phi_{\rm s}-\phi_{\rm d}$ to a fractional accuracy of $10^{-2}$.

\subsubsection{Strong interactions \label{sec:strong_interactions}}

The above treatment assumed that interaction energies could be computed perturbatively using the non-interacting ground state consisting of both $e$ and $g$ atom in the motional ground state of the magic-wavelength tweezer potential.  More generally, the interaction energy for triplet and singlet electronic states in a magic wavelength tweezer can be computed exactly in the pseudopotential approximation (as long as very small anharmonic corrections to the trapping potential are ignored) \cite{Busch1998}. Assuming isotropic confinement along all three directions, achievable by retro-reflecting part of the tweezer laser intensity for increased axial confinement, it is possible to compute $U^\pm_{eg}=E^{\pm}-\frac{3}{2}\hbar\omega_{\rm tw}$, with $E^{\pm}$ satisfying the equation:
\begin{equation}
\label{eq_}
\sqrt{2}\, \frac{\Gamma(-E^{\pm}/2 + 3/4)}{\Gamma(-E^{\pm}/2 + 1/4) }= \frac{1}{a_{eg}^\pm/a_{\rm tw}}.
\end{equation}
Here $\Gamma$ is the Gamma function, $(a_{eg}^-,a_{eg}^+)=(219.5, 1878)\,a_0$ for $^{173}$Yb \cite{Hofer2015} and $(a_{eg}^-,a_{eg}^+)=(69.1, 160)\,a_0$ for $^{87}$Sr \cite{Goban2018}, with $a_0$ being the Bohr radius.
With reasonable laser powers at magic wavelengths (see Tables \ref{table_1}-\ref{table_2} in Appendix \ref{app_E}),
it is possible to achieve trap depths of $\omega_{\rm tw}/2\pi\simeq 100\,\rm kHz$ for the individual tweezers and $\omega_{\rm g}/2\pi\simeq 170\, \rm kHz$ for the transport tweezers. 

In the case of $^{87}$Sr, the typical interaction energies at the end of the transport tweezer ramp down will be much lower ($V_{\rm ex}/h=6\,{\rm kHz},\, V_{\mathrm d}/h=15~\mbox{kHz}$), indicating that a perturbative treatment is well justified. In the case of $^{173}$Yb, the electronic triplet interaction nearly saturates to the vibrational gap $\hbar \omega_{\rm tw}$ \cite{Busch1998}, while the singlet interaction is well within the perturbative regime for $\omega_{\rm tw}$ up to 150 kHz. The singlet and triplet two-body motional wavefunctions will therefore differ significantly, and the exchange dynamics induced by suddenly turning off the OS beam will lead to a complicated interplay of spin and motional dynamics, invalidating the simple Hamiltonian (\ref{eq_H_eg}) used in the previous section.

There are several potential routes to avoiding these complexities. One option is to keep the transport tweezer on during the exchange gate, and use it to maintain a sufficiently small overlap between $\e$ and $\g$ atoms for a perturbative treatment of the interactions to be justified. Alternatively, one could avoid using an OS beam altogether; in this case the singlet-triplet basis diagonalizes the interaction for all tweezer configurations, and a gate will result from the relative phase difference accumulated in the singlet/triplet channels during the merger.  The second approach can in principle yield an extremely fast gate for $^{173}$Yb, but suffers from a sensitivity to pointing fluctuations during the transport.

\section{Register initialization}\label{sec_reg_in}

The 1D tweezer array can be initialized with one atom per trap starting from a low-temperature, narrow-line magneto optical trap on the $^1S_0\rightarrow{^3P_1}$ transition. Light-assisted collisions will ensure to have at least half of the tweezers occupied by one single atom \cite{Endres2016, Yamamoto2016}. 
The singly trapped atoms can be detected with low atom losses (see section \ref{sec_readout} for details), and rearranged in a 1D array with uniform spacing. The rearranging can be achieved through the dynamically control of acousto-optical beam deflectors \cite{Endres2016,Labuhn2016}. The atom-atom spacing $d$ is chosen so that the cross-talk between separate tweezers is negligible, namely the individual minima are well defined, while at the same time the distance for transport is minimized. Both conditions are satisfied for $d\simeq 3 \lambda_m$, where $\lambda_m$ is the magic wavelength in consideration.
After rearranging the individual tweezers, optical pumping to the $\ket{g\!\downarrow}$ state can be efficiently performed by resonantly addressing the $^1S_0\rightarrow{^3P_1}$ transition with $\sigma^{-}$ polarized light.

\begin{figure}[t!]
\centering
\includegraphics[width=\columnwidth]{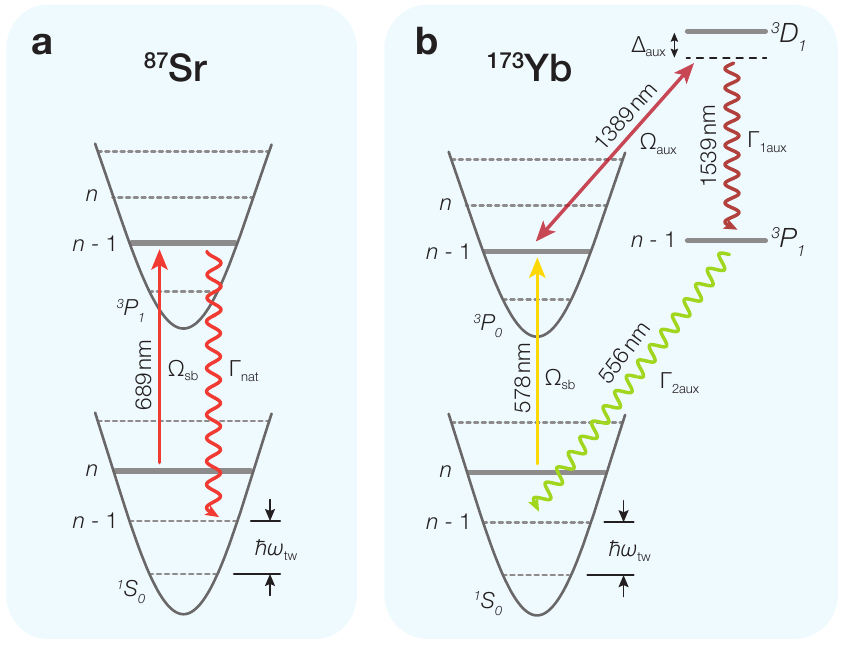}
\caption{{\bf Resolved-sideband cooling}: Ground state cooling schemes for $^{87}$Sr (\textbf{a}) and $^{173}$Yb (\textbf{b}). For Yb, in order to enhance the cooling rate on the ${}^1S_1 \rightarrow {}^3P_0$ transition, the ${^3}P_0$ level is quenched by a re-pumping laser set to a detuning $\Delta_\textrm{aux}$ from the ${^3}P_0 \rightarrow {^3}D_1$ transition.}
\label{fig_sideband}
\end{figure}

Subsequently, the atoms can be cooled to the tweezer motional ground state by using cycles of optical pumping and resolved-sideband cooling \cite{Nemitz2016, Brown2017}. 
This can be achieved in both atomic species, considering $\omega_{\rm tw}/2\pi \sim 100\,\rm kHz$. For $^{87}$Sr, the $^1S_0\rightarrow{^3}P_1$ narrow transition with $\Gamma_{\rm nat}/2\pi=7.4\,\rm kHz$ is well within the sideband-resolved regime. \fsB{This provides a cooling rate of 10 quanta/ms with $\Omega_{\rm sb}/2\pi\simeq 20\,\rm kHz$, yielding a final ground-state population $n_0 \simeq 0.99$}. 
In the case of $^{173}$Yb, the $^1S_0\rightarrow{^3}P_1$ transition has a linewidth $\Gamma/2\pi=182\,\mbox{kHz}$ and can only yield an average population of a few motional quanta. The final step of ground-state cooling can be performed exploiting the clock transition \cite{Curtis2001, Nemitz2016, Brown2017}, broadened by quenching the $^3P_0$ level with the application of $^3P_0\rightarrow {^3D}_1$ repumping light at 1389\,nm. This results in an effective tunable scattering rate \cite{Leibfried2003}:
\begin{equation}
\label{eq_Gamma_eff}
\Gamma_{\rm eff}=\frac{\Omega_{\rm aux}^2}{(\Gamma_{\rm nat}+\Gamma_{\rm 1aux})^2+4\Delta_{\rm aux}^2}\Gamma_{\rm aux},
\end{equation}
\noindent
where $\gamma_{\rm nat}$ is the linewidth of the clock line, $\Omega_{\rm aux}$ and $\Delta_{\rm aux}$ are the Rabi frequency and the detuning of the $^3P_0\rightarrow {^3D}_1$ rempumper, and $\Gamma_{\rm aux}^{-1}=\Gamma_{1\rm aux}^{-1}+\Gamma_{2\rm aux}^{-1}$, where $\Gamma_{1\rm aux}^{-1}=1.06 \,\mu\rm s $ and $\Gamma_{2\rm aux}^{-1}=300\,\rm ns$ are the lifetimes of the ${^3D}_1$ and ${^3}P_1$ states, respectively (see Fig.~\ref{fig_sideband}). The atoms pumped in the $^3P_1$ state decay to the ground state $^1S_0$, emitting 556\,nm photons, which can also be used for detection on the CCD. Atom imaging during sideband cooling will be limited by the relatively long lifetime of the $^{3}P_1$ state and the small branching ratio on the $^3P_2$ state (26:1000). However, a big advantage is that the 556\,nm light can be filtered out with respect to other wavelengths involved in the process ($^3P_0\rightarrow{^3}D_1$ at 1389\,nm and $^3D_1\rightarrow{^3}P_1$ at 1539\,nm).
With $\Omega_{\rm aux}/2\pi=\Delta_{\rm aux}/2\pi=10\,$MHz and $\Omega_{\rm sb}/2\pi= 20\,$kHz, it is possible \fsB{to achieve a ground-state population $n_0 \simeq 0.98$} through a cooling rate of $\sim 0.7$ quanta/ms. For both species, ground-state cooling is limited by the heating rate associated with the deconfinement of the $^3P_1$ state in  magic-wavelength tweezers. This issue can be solved by performing sideband cooling within tweezers tuned at the magic wavelength between $^1S_0$ and $^3P_1$.

\fsB{In a realistic experimental scenario, one or more motional excitations may persist in the array due to imperfect ground-state cooling. The exchange energy associated with atoms occupying vibrationally excited states is significantly reduced, considerably affecting the fidelity of two-qubit gates (see Appendix~\ref{app_D}). However, motionally excited atoms can be selectively removed after sideband cooling through imaging of the $^3P_0$ state. Following a final red-sideband clock pulse, $^3P_0$ state detection can be achieved using $^3D_1$ or $^3S_1$ rempumping lasers. This allows to directly identify and drop all excited-state atoms participating to the red sideband, and to suitably rearrange the array, further extracting entropy from the system.}

Following ground state cooling, the spin-polarized $\g$-state array needs to be initialized into a staggered $\e-\g$ configuration (see Fig.~\ref{fig_1}). This can be realized by single-tweezer addressing with a clock beam resonant with the clock $^1S_0\rightarrow{^3}P_0$ transition, required also to perform individual single-qubit gates. To avoid cross-talking with the neighboring $g$ atoms on every other site, it is also possible to use the OS or transport tweezer beams to selectively shift the targeted atom in resonance.

\section{Read-out}\label{sec_readout}

In order to read out the final state of the full quantum register, we first perform a mapping of the nuclear qubit state of the atoms in each tweezer to the electronic state. For odd (even) tweezers, atoms are found in a linear nuclear spin superposition of $\ket{g}$ ($\ket{e}$). The mapping is performed by applying a clock $\pi$-pulse with $\pi$-polarization on every atom, using a frequency resonant with the $\ket{g,\uparrow} \leftrightarrow \ket{e,\uparrow}$ transition. In this way, the state $\ket{\!\!\uparrow}$ is mapped to the $\ket{e}$ ($\ket{g}$) state for atoms in odd (even) traps, whereas the $\ket{\!\!\downarrow}$-state corresponds to the $\ket{g}$ ($\ket{e}$) state for atoms in odd (even) traps. At this point, $\ket{g}$-state atom detection is equivalent to nuclear spin detection, with bright (dark) atoms in odd traps associated to the $\ket{\!\!\downarrow}$ ($\ket{\!\!\uparrow}$) state, and vice-versa in even traps. 

Single-atom resolved $\ket{g}$-state detection is performed using fluorescence imaging on the strong $\ket{g}\rightarrow\ket{d}$ transition, with $\ket{d}=\ket{^1P_1}$, which provides a conveniently high scattering rate $\Gamma_g \approx 2 \times 10^8$ photons/s. This allows for a fast single-atom resolved detection of the $\g$ population in each tweezer, since several hundreds of photons can be scattered during an exposure time on the order of 100$\,\mu$s \cite{Miranda2014}. Using currently available EMCCD sensors for fluorescent photon collection, few hundreds of counts suffice to discriminate traps where atoms in the bright state are present from empty ones.  During 100$\,\mu$s of fluorescence imaging, atoms will be heated to a few hundreds $\mu$K, a value still well below the effective trap depth experienced by atoms during the excitation process $(V_m^{g} + V_m^d)/2 \sim 1$\,mK.

If larger fluorescence signals are required for increasing the detection fidelity, the fluorescence collection time can be greatly extended by additionally applying sub-Doppler molasses cooling or sideband cooling operating on the narrow $^1S_0\rightarrow{^3}P_1$ intercombination transition, similar to what already demonstrated in a $^{174}$Yb quantum gas microscope \cite{Yamamoto2016}. This would allow to collect more than 1000 photons per atom without atom losses from the tweezers. 

Ideally, the $\ket{g}$-state detection does not perturb the $\ket{e}$-state atoms, and therefore one can apply a subsequent clock $\pi$-pulse to invert the clock state populations. Therefore, it is possible to measure the $\ket{e}$-state population as well, thereby improving statistics and isolating preparation/read-out errors.





\section{Conclusions and Outlook}

In this work we have laid out a new platform for quantum information processing with neutral fermionic AEL atoms via a realistic architecture, that appears entirely within reach of state-of-art atomic physics experiments. 

%

\gp{
The peculiar properties of AEL atoms allow for high resolution in both the spatial and spectral domain. In particular, the synergy between a robust nuclear qubit and an auxiliary optical qubit yields a high flexibility, together with the use of off-resonant state-dependent tweezer potentials to perform efficient single-atom transport and rearrangements within the array. The presence of a highly coherent spin-exchange coupling enables fast, optically gated entangling operations, which are largely insensitive to beam pointing and intensity fluctuations. 
In addition, we have devised schemes for state preparation, single-qubit coherent manipulation, and detection at the single-atom level. 
}
We also explored locally adiabatic protocols for moving and merging individual atoms in the weakly-interacting (perturbative) case. The extensions of such protocols to the case of strongly interacting atoms \cite{Fogarty2018} remains an interesting open question for future studies. 

Such a versatile platform offers exciting prospects for both quantum computing and digital quantum simulation, and it can be straightforwardly extended to two-dimensional arrays.
Considering that the entangling gates on different pairs can be performed in parallel with virtually no crosstalks between distinct qubits, this platform is ideal for the realization of genuine multi-partite entangled states with $\mathcal{O}(N)$ operations. In particular, the reliable production of cluster states \cite{Briegel2001, Briegel2009, Mandel2003} would enable the implementation of one-way quantum computing algorithms \cite{Raussendorf2003, Briegel2009}, where local measurements can be achieved through the scheme proposed in Ref. \cite{Daley2008}. Additionally, the precise control of both internal and external degrees of freedom enabled by this architecture could be used for investigating RKKY and spin-orbital interactions in minimal realizations of the Kondo lattice and Kugel-Khomskii models \cite{Gorshkov2010, Foss-Feig2010}. \fsB{In this context, the addition of a tunable-wavelength optical lattice \cite{Ono2018} would provide the required flexibility for realizing two-orbital lattice Hamiltonians on demand}.

\fsB{An appealing modification of this architecture can be envisioned using Rydberg interactions for performing two-qubit gates. In this case, all atoms in the array would be initialized in state $\ket{g}$. A two-qubit operation could be performed by exciting two neighbouring atoms to state $\ket{e}$ in a spin-selective fashion, using the single-qubit operations described in Section~\ref{sec_gates}. A global Rydberg excitation pulse acting on the $e$ atoms would then apply a state-dependent phase, which after transferring the atoms back to state $\ket{g}$ would result in a phase gate on the nuclear qubits. Rydberg $s$-states can indeed be reached from the $^3P_0$ state through a single-photon transition, leading to increased single-photon Rydberg Rabi frequencies with respect to alkali atoms, therefore allowing for fast operations and mitigating decoherence.}


\section{Acknowledgments}
The authors would like to thank Marcello Dalmonte, Adam Kaufman, Leonardo Fallani and Jeff Thompson for useful discussions and suggestions. G.P. is supported by the ARO and AFOSR Atomic and Molecular Physics Programs, the AFOSR MURI on Quantum Measurement and Verification, the IARPA LogiQ program, and the NSF Physics Frontier Center at JQI. F.S.~acknowledges support from the European Union H2020 Marie Sk\l{}odowska-Curie program (grant no.~705269).

\smallskip
\textit{Note added} - After submission of this manuscript, experimental work has been reported on sideband cooling and single-atom imaging of two-electron atoms in optical tweezer arrays \cite{Cooper2018, Norcia2018, Saskin2018, Covey2018}.

\appendix 

\section{Motional ground state fidelity during transport}\label{app_A}

Nearest neighbor atoms are separated by a distance much greater than the extent of their ground state wavefunctions, and must be transported into the same location to initiate an exchange gate.  This transport can easily pose a bottleneck in the gate speed, and therefore it is important to understand how quickly it can be achieved without causing appreciable motional excitations.  Here, we calculate the persistence probability $\mathcal{F}$ for an atom initially in the motional ground state of a tweezer to remain in the final (instantaneous) motional ground state after a time-dependent manipulation of the tweezer potential.

During the transport of a $g$-atom, the potential is a sum of the magic wavelength potential and the transport tweezer (which only traps the $g$ atom).  Calling the total potential $\hat{V}(t)$ and working in the harmonic approximation $\hat{V}(t)\approx \frac{1}{2}m \omega(t)^2[\hat{x}-x_{\rm c}(t)]^2$, the Hamiltonian for the $g$ atom can be written
\begin{align}
\hat{H}(t)=\Omega(t)\hat{a}_0^{\dagger}\hat{a}_0+\frac{\eta(t)}{2}(\hat{a}_0^2+\hat{a}_0^{\dagger 2})-F(t)(\hat{a}_0+\hat{a}_0^{\dagger})+c(t).
\end{align}
Here $\hat{a}_0$($\hat{a}_0^{\dagger}$) is the anihilation(creation) operator for excitations of the initial ($t=0$) Hamiltian, with corresponding frequency $\omega_0$ and harmonic-oscillator width $x_0=\sqrt{\hbar/(m\omega_0)}$. The time-dependent energy shift $c(t)$ is of no consequence in what follows, and
\begin{align}
\Omega(t)&=\frac{\hbar\omega_0}{2}\big([\omega(t)/\omega_0]^2+1\big),\\
\eta(t)&=\frac{\hbar\omega_0}{2}\big([\omega(t)/\omega_0]^2-1\big),\\
F(t)&=\frac{\hbar\omega(t)^2}{\sqrt{2}\omega_0}\frac{x_{\rm c}(t)}{x_0}.
\end{align}

There are many approaches to computing dynamics in time-dependent harmonic potentials, for example the invariant method of Lewis and Riesenfeld \cite{doi:10.1063/1.1664991}.  Here we follow a less technical route, and begin by writing the Heisenberg picture anihilation operator as $\hat{a}(t)=u(t)\hat{a}_0+v(t)\hat{a}^{\dagger}_0+z(t)$.  That an expansion in $a_0$ and $\hat{a}^{\dagger}_0$ truncates exactly at first-order is a unique property of quadratic Hamiltonians, and underlies the solvability of the problem.  Now we take the time derivative of $\hat{a}(t)$ as
\begin{align}
\frac{d\hat{a}(t)}{dt}&=i \big(\hat{U}^{\dagger}[\hat{H}(t),\hat{a}_{0}]\hat{U}\big)\nonumber\\
&=-i\big(\Omega(t)\hat{a}(t)+\eta(t)\hat{a}^{\dagger}(t)-F(t)\big).
\label{eq:time_deriv}
\end{align}
Inserting the expansion for $\hat{a}(t)$ back into Eq.\ (\ref{eq:time_deriv}), and then equating operator coefficients when comparing to the explicit derivative $d\hat{a}(t)/dt=\hat{a}_0 \dot{u}(t)+\hat{a}^{\dagger}_0 \dot{v}(t)+\dot{z}(t)$, we obtain equations of motion for the expansion coefficients
\begin{align}
\label{eq:eqsom}
i\dot{u}(t)&= \Omega(t)u(t)+\eta(t)\bar{v}(t),\\
i\dot{v}(t)&= \Omega(t)v(t)+\eta(t)\bar{u}(t),\\
i\dot{z}(t)&= \Omega(t)z(t)+\eta(t)\bar{z}(t)-F(t).
\end{align}

A convenient explicit representation of the time evolution operator can be obtained by writing 
\begin{align}
\hat{U}(t)=\hat{D}\big(\alpha(t)\big)\hat{S}\big(\zeta(t)\big)\hat{R}\big(\theta(t)\big),
\end{align}
where $\hat{D}$, $\hat{S}$, and $\hat{R}$ are displacement, squeezing, and rotation operators, respectively. Using standard identities to write $\hat{U}^{\dagger}(t)\hat{a}_0 \hat{U}(t)$ in terms of $\alpha(t)$, $\zeta(t)$, and $\theta(t)$, and equating operator coefficients with the expansion of $\hat{a}(t)$ in terms of $u(t)$, $v(t)$, and $z(t)$, we obtain
\begin{align}
\label{eq:conversion}
\theta(t)&= - \arg u(t), \\
\zeta(t)&= \cosh^{-1}|u(t)|e^{i(\pi-\arg u(t)-\arg v(t))},\\
\alpha(t)&=z(t).
\end{align}

The probability for an atom initially in the ground state of $\hat{H}(0)$ to be in the instantaneous ground state of $\hat{H}(t)$ at time $t$ is given by
\begin{align}
\label{eq:prob0to0}
P_{0\rightarrow 0}(t)=|\bra{\Psi_0(t)}\hat{U}(t)\ket{\Psi_0(0)}|^2,
\end{align}
where $\ket{\Psi_0(t)}$ is the instantaneous ground state of $\hat{H}(t)$.  Using the following compact notation for a displaced and squeezed coherent state
\begin{align}
\ket{\zeta_,\alpha}\equiv \hat{D}(\alpha)\hat{S}(\zeta)\ket{\Psi_{0}(0)},
\end{align}
it is straightforward to show that
\begin{align}
\ket{\Psi_0(t)}=\ket{\zeta_{\rm inst}(t),\alpha_{\rm inst}(t)},
\end{align}
where $\alpha_{\rm inst}(t)=\frac{x_{\rm c}(t)}{\sqrt{2}x_0}$ and $\zeta_{\rm inst}(t)=\frac{1}{2}\log(\omega(t)/\omega(0))$. Inserting the expression for $\hat{U}(t)$ and $\ket{\Psi_0(t)}$ into Eq.\ (\ref{eq:prob0to0}), and noting that the rotation operator acts trivially on a harmonic-oscillator ground state, we find
\begin{align}
\label{eq:p_fin}
P_{0\rightarrow 0}(t)=|\bra{\zeta_{\rm inst}(t),\alpha_{\rm inst}(t)}\zeta(t),\alpha(t)\rangle|^2.
\end{align}
With $\zeta(t)$ and $\alpha(t)$ obtained numerically, the right-hand side of Eq.\ (\ref{eq:p_fin}) can be evaluated using the following result from Ref.\ \cite{PhysRevA.54.5378},
\begin{align}
|\bra{z_1,\alpha_1}z_2,\alpha_2\rangle|^2=\frac{e^{\Re(\eta_{21}\bar\eta_{12}/2\sigma)}}{|\sigma|},
\end{align}
where
\begin{align}
\sigma&=\cosh|z_1|\cosh|z_2|-e^{i(\arg z_1-\arg z_2)}\sinh|z_1|\sinh|z_2|,\nonumber\\
\eta_{jk}&=(\alpha_j-\alpha_k)\cosh|z_j|-e^{i\arg z_j}(\bar{\alpha}_j-\bar{\alpha}_k)\sinh|z_j|.\nonumber\\[5pt]
&
\end{align}

\begin{figure}[t!]
\centering
\includegraphics[width=1.\columnwidth]{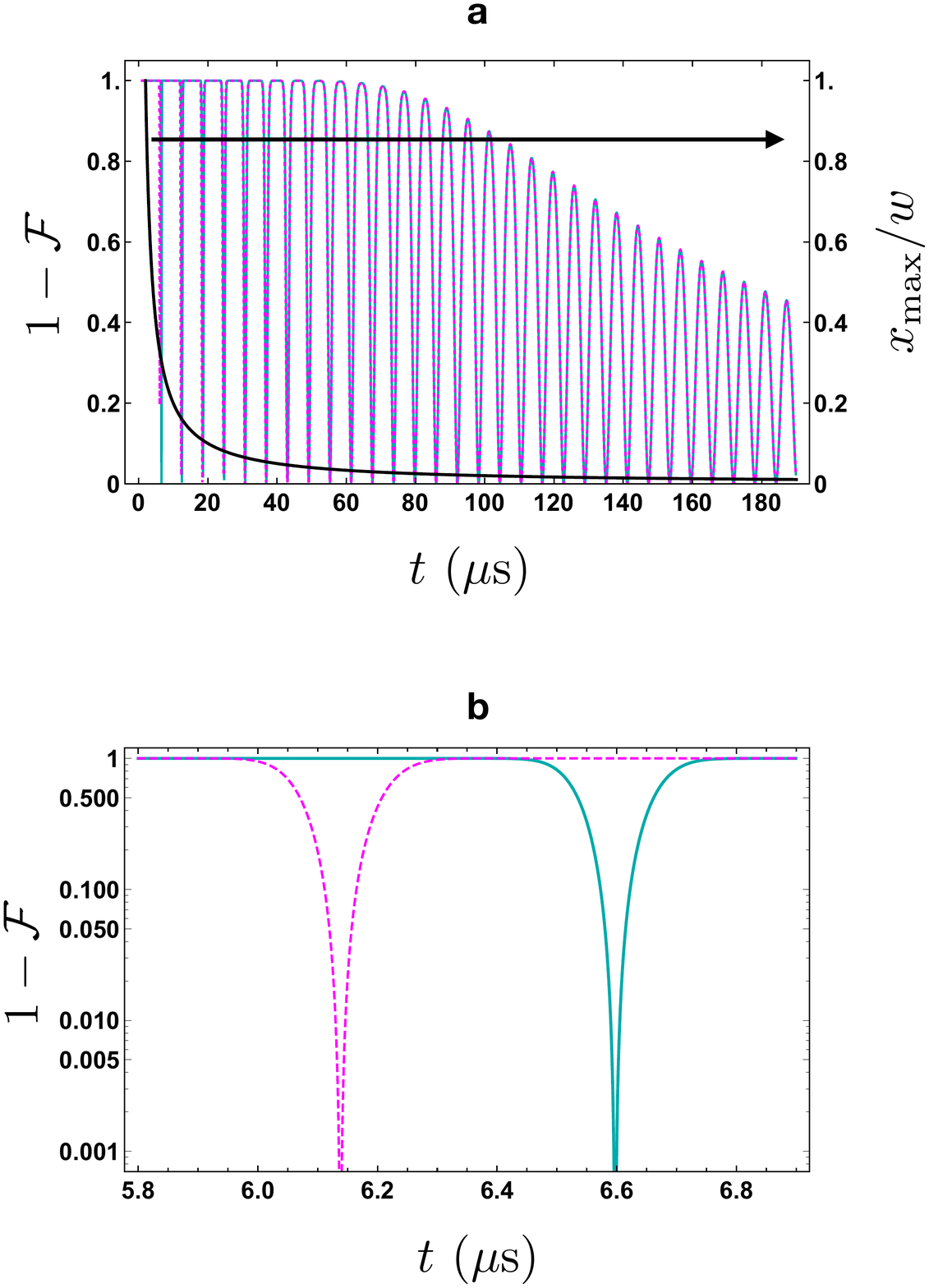}
\caption{{\bf Transport corrections due to the Gaussian tweezer potential}. All parameters are the same as in Fig.\ \ref{fig_4} of the manuscript.  (\textbf{a}) Purple dashed line is a reproduction of Fig.\ \ref{fig_4} of the manuscript.  The blue solid line (mostly concealed behind the dashed line) is the infidelity calculated classically using the full Gaussian tweezer potential (rather than in the harmonic approximation).  The black curve shows the maximum excursion (in units of the beam waist) of a trajectory with duration $t$.  (\textbf{b}) Zoom in of (a) around the first infidelity dip.  Note that the exact tweezer potential leads to a slight delay of the infidelity dip, as the Gaussian potential softens (slowing oscillations) away from the beam center.
}
\label{fig_s1}
\end{figure}

For fast (diabatic) transport, the atom may make transient excursions into the non-harmonic wings of the Gaussian tweezer beam.  For example, the first dip of the infidelity in Fig.\ \ref{fig_4}(b) corresponds to a transport protocol in which the atom makes excursions of radius $\sim 30\%$ of the beam waist.  A fully classical (i.e. ignoring squeezing) calculation of the transport dynamics in the true Gaussian potential suggests that the effect of the higher-order corrections to the harmonic potential are relatively minor, and primarily involve a slight delay of the infidelity dips in the diabatic regime, as shown in Fig.\ \ref{fig_s1} (see in particular Fig.\ \ref{fig_s1}(b) for the most diabatic transport time considered). As seen in Fig.\ \ref{fig_s1}(a), for transport on timescales of tens of $\mu$s the effect of classical dynamics in a true Gaussian potential and in the harmonic approximation are nearly identical. The black curve shows the maximum extent of a trajectory, $x_{\rm max}$, as a function of the trajectory duration. By $\sim20\,\mu{\rm s}$, the maximum deviations are only $\sim10\%$ the beam waist, implying corrections to the harmonic potential at the $\sim1\%$ level.

\section{Optical switch and associated scattering rates}\label{app_B}

\begin{figure}[t!]
\centering
\includegraphics[width=1.\columnwidth]{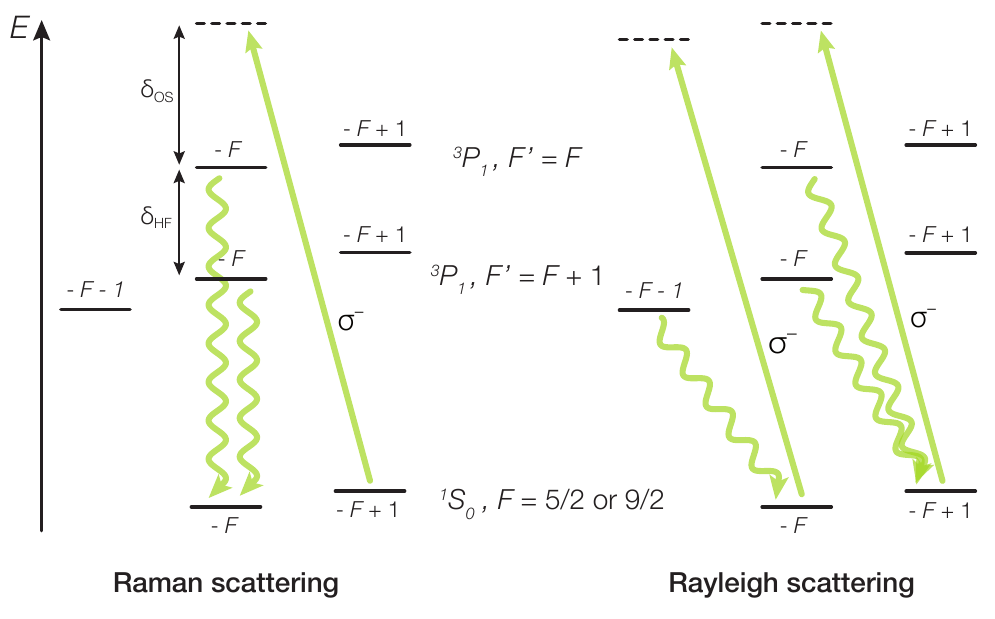}
\caption{{\bf OS beam-induced decoherence processes:} scattering paths through OS laser absorption and spontaneous emission. The $\sigma^-$ polarized light causes off-resonant optical pumping of states $\ket{2}$ and $\ket{4}$ into states $\ket{1}$ and $\ket{3}$, featuring a $\ket{g\downarrow}$ atom. 
}
\label{fig_6}
\end{figure}

The optical switch (OS) principle exploits the advantageous large ratio between the hyperfine splitting $\delta_\text{HF}$ of the ${^3}P_1$ state and the small natural linewidth of the $^1S_0\rightarrow{^3}P_1$ transition, typical of AEL atoms, to efficiently generate a spin-dependent light shift for $\g$ atoms \cite{Stellmer2011, Taie2010}. Indeed $\delta_\text{HF}$, i.e.~the energy difference between the $F' = F$ and $F' = F+1$ manifolds of the ${^3}P_1$ state, is on the order of few $h \times$\,GHz, for both $^{173}$Yb or $^{87}$Sr.

The OS beam parameters, namely polarization, intensity and detuning, determine  the strength of the spin-dependent light shift and the photon scattering rate for $\g$ atoms. A $\sigma^-$-polarized beam with detuning $\delta_\text{OS} = 45\,$GHz or $\delta_\text{OS} = 5\,$GHz from the $^1S_0\rightarrow{^3}P_1$ transition is chosen for $^{173}$Yb or $^{87}$Sr, respectively. Such values represent a trade-off between the required beam intensity and the induced spontaneous photon scattering, given a differential light shift of the order of $V_\text{OS}/h \approx 1$\,MHz between the $m_F = - F$ and $m_F = - F + 1$ states.

In order to compute the decoherence rates associated with the near-resonant OS beam, we considered the processes that effectively measure the qubit state \cite{Uys2010}, namely:
\begin{eqnarray}
\label{eq_scattering_rates}
\Gamma_{\uparrow\downarrow}&=& \Omega^2\Gamma \left(\sum_{F'} \frac{C_{F'\downarrow}C_{F'\uparrow}}{2\delta_{F'}}\right)^2,\\
\Gamma_{el}&=& \Omega^2\Gamma 
\left(\sum_{F'} \frac{|C_{F'\downarrow}|^2}{2\delta_{F'}}
-\frac{|C_{F'\uparrow}|^2}{2\delta_{F'}}
\right)^2,
\end{eqnarray}
where $2\Omega^2/(2J'+1)^2\Gamma^2=I/I_s$, with $\Gamma=2\pi\times 182\,(7.4) \,\rm kHz$ being the natural linewidth of the $3P_{J'=1}$ state $^{173}$Yb ($^{87}$Sr), $I_s$ the saturation intensity. $C_{F'\sigma}$ with $\sigma=\uparrow,\downarrow$ are the Clebsch-Gordan coefficients connecting the spin states $\sigma$ with the excited states $^3P_1F'$ with the respective detuning $\delta_{F'}$.
It shall be noticed that the choice of the $\sigma^-$-polarized light and the large detuning from the $^1S_0\rightarrow{^3P}_1$ transition reduce significantly both the Raman and Rayleigh scattering rates. It is possible to achieve $\Gamma_{\uparrow\downarrow}/(2\pi)\simeq 5\times10^{-7}\,V_\textrm{OS}/h$ and $\Gamma_{el}/2\pi\simeq 1\times 10^{-7}\,V_\textrm{OS}/h$, by choosing $\delta_\text{OS} =45\, (5)\,{\rm GHz}$ from the ${^3P}_1F'=F+1$ for $^{173}$Yb ($^{87}$Sr).

In order to estimate the motional heating $\eta^2\Gamma_\sigma$, the total scattering rate $\Gamma_\sigma$ for the spin state $\sigma=\uparrow,\downarrow$ needs to be take into account:
\begin{equation}
\label{motional}
\Gamma_\sigma=\Omega^2\Gamma 
\left(\sum_{F'} \frac{|C_{F'\sigma}|^2}{2\delta_{F'}}
\right)^2.
\end{equation}

\section{Decoherence dynamics}\label{app_master}\label{app_C}
The decoherence associated with scattering processes and the unitary errors induced by the OS beam can be accounted for by using a master equation to solve exactly the dynamics in presence of decoherence:
\begin{equation}
\label{eq_master}
\partial_t \rho=\frac{i}{\hbar} [H,\rho] + \mathcal{L}(\rho),
\vspace*{4pt}
\end{equation}
where $\rho$ is the density matrix and $\mathcal{L}$ is the Lindbladian superoperator, which can be defined as:
\vspace*{4pt}
\begin{widetext}
\begin{equation}
\small
\label{eq_Limblad}
\mathcal{L}(\rho)=
\begin{pmatrix}
\Gamma_{\uparrow\downarrow} \rho_{22}-\eta^2\Gamma_{\downarrow}\rho_{11} & -(\Gamma_{\uparrow\downarrow}+\Gamma_{el}+\eta^2\Gamma_{\rm tot}/2 \,\rho_{12} 	& 0 & 0\\
  -(\Gamma_{\uparrow\downarrow}+\Gamma_{el}+\eta^2\Gamma_{\rm tot})/2 \,\rho_{21}  & -\Gamma_{\uparrow\downarrow} \, \rho_{22} - \eta^2\Gamma_{\uparrow}\,\rho_{22} & -(\Gamma_{\uparrow\downarrow}+\Gamma_{el}+\eta^2\Gamma_{\rm tot})/2 \,\rho_{23} & 0\\
  0   &-(\Gamma_{\uparrow\downarrow}+\Gamma_{el}+\eta^2\Gamma_{\rm tot})/2 \,\rho_{32}  &\Gamma_{\uparrow\downarrow} \rho_{44} - \eta^2\Gamma_{\downarrow}\,\rho_{33} & -(\Gamma_{\uparrow\downarrow}+\Gamma_{el}+\eta^2\Gamma_{\rm tot})/2 \,\rho_{34}\\
  0   &0 & -(\Gamma_{\uparrow\downarrow}+\Gamma_{el}+\eta^2\Gamma_{\rm tot})/2 \,\rho_{43}& -\Gamma_{\uparrow\downarrow} \rho_{44} - \eta^2\Gamma_{\uparrow}\,\rho_{44} \nonumber \\[2mm]
\end{pmatrix},
\normalsize
\end{equation}
\end{widetext}
where the states $\ket{1},\ket{2},\ket{3},\ket{4}$ follow from Eq. (\ref{eq_basis}). The $\sigma^-$-polarized optical switch beam will cause Raman scattering events affecting only two states ($\ket{2}$ and $\ket{4}$) out of four, namely the ones including a $\ket{g\!\uparrow}$ atom (see Fig. \ref{fig_6}b). The motional heating is modeled by inducing a non trace-preserving decay proportional to $\eta^2\Gamma_{\sigma}$, where $\Gamma_{\rm tot}=\Gamma_{\uparrow}+\Gamma_{\downarrow}$.

\vspace{8pt}

\section{Impact of imperfect cooling}\label{app_D}
Here we extend the treatment of the previous section to the case of imperfect ground-state cooling. We restrict to the case of one single motional excitation shared by one of the two atoms undergoing the collisional gate.  
We consider contributions arising from the symmetric and anti-symmetric vibrational states $\ket{01}_s$ and $\ket{01}_a$ defined as:
\begin{eqnarray}
\label{eq_vibstates}
\ket{01}_s&=&\frac{\ket{01}+\ket{10}}{\sqrt{2}}\nonumber\\
\ket{01}_a&=&\frac{\ket{01}-\ket{10}}{\sqrt{2}}
\end{eqnarray}
Therefore recalling Hamiltonian (\ref{eq_H_eg}), we can write the full Hamiltonian as
\begin{equation}
H=H_{eg}\otimes \mathbb{1}_{0} + H^{(s)}_{eg}\otimes \mathbb{1}_{s} + H^{(a)}_{eg}\otimes \mathbb{1}_a
\end{equation}
where $H^{(s)}_{eg}$ and $H^{(a)}_{eg}$ are the interaction Hamiltonians associated to the two states defined in Eq. (\ref{eq_vibstates}) and $\mathbb{1}_i$ is the identity matrix in the motional subspace $i=0,a,s$. On the one hand, $H^{(s)}_{eg}$ features interaction energies very close to ground state Hamiltonian $H_{eg}$, because the antisymmetrization compensates almost exactly the reduced integral overlap between the ground and first motional excited state wavefunctions (this compensation is exact for harmonic oscillator wavefunctions). On the other hand, $H^{(a)}_{eg}$ features no contact interactions at all, since the spatial wavefunction is antisymmetric. 

\begin{figure}[b!]
\centering
\includegraphics[width=0.9\columnwidth]{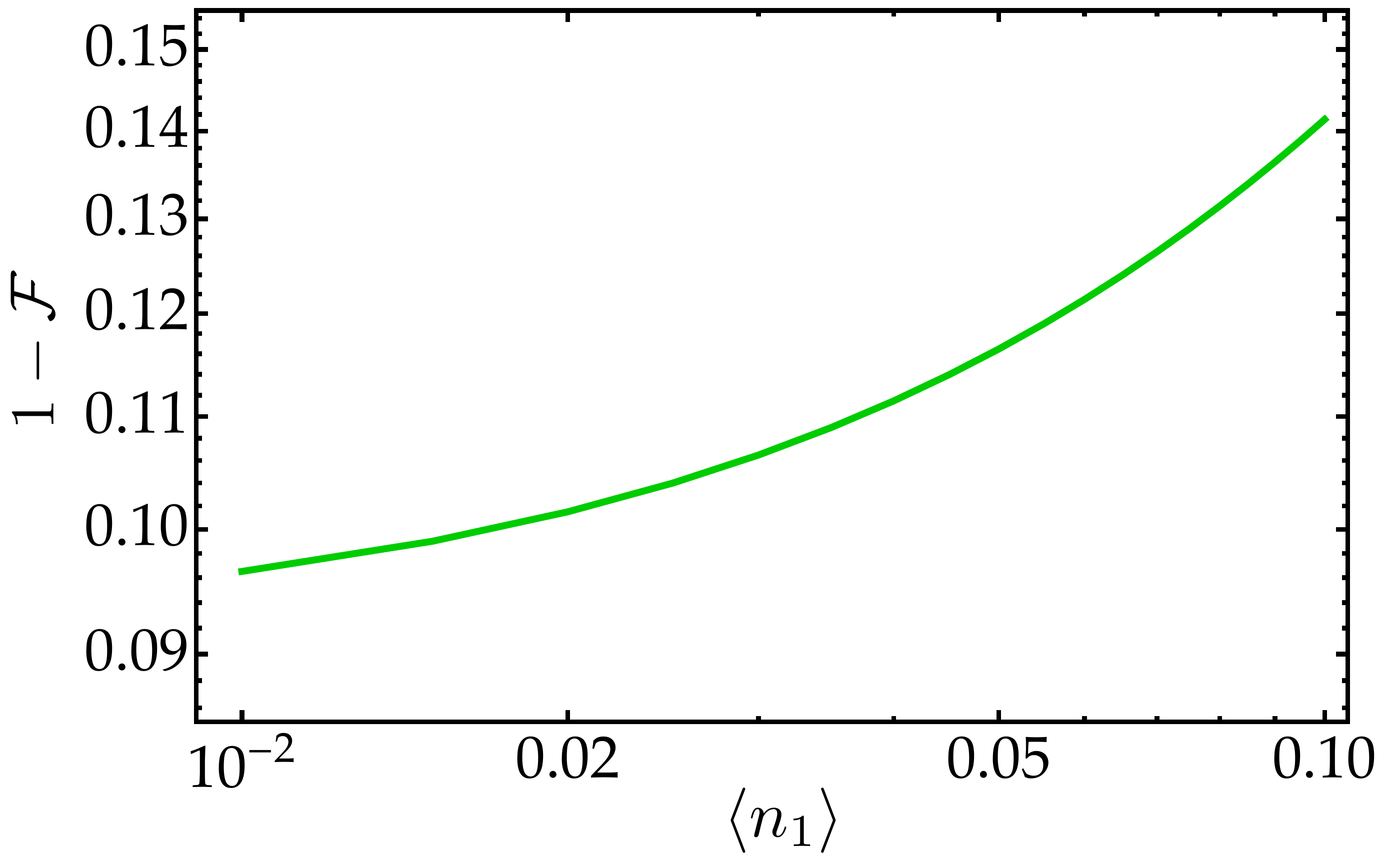}
\caption{ \fsB{
{\bf Fidelity for imperfect ground-state cooling.} Error contribution to the two-qubit gate fidelity of a population $\langle n_1\rangle$ in the first vibrationally excited state. The solid green line shows the error $1-\mathcal{F}$ after 40 repeated two-qubit $\rm SWAP$ gates as a function of $\langle n_1\rangle$. The fidelity is calculated by computing the overlap with the desired spin state, while tracing over the motional degrees of freedom.}}
\label{fig7}
\end{figure}

The Lindbladian superoperator acting on the density matrix can be generalized as:
\begin{equation}
\mathcal{L}(\rho)=\mathcal{L}_{0}(\rho) \otimes \mathbb{1}_{s} \otimes \mathbb{1}_{a} + \mathbb{1}_0 \otimes \mathcal{L}_{s}(\rho)\otimes \mathbb{1}_a + \mathbb{1}_0 \otimes \mathbb{1}_s \otimes \mathcal{L}_{a}(\rho){ \over }, 
\vspace*{4pt}
\end{equation}
where $\mathcal{L}_{0}(\rho)$ accounts for the heating and dephasing processes for the atoms pair in the ground state defined in Eq. (\ref{eq_Limblad}), while $\mathcal{L}_{s,a}$ accounts for dissipative processes in the motionally excited subspace, taking into account both the spontaneous emission processes, identical to that of atoms in the motional ground state, and the incoherent pumping from the ground state into the motional excited states caused by motional heating (see section \ref{sec_2qubit}). Assuming that the initial state is a statistical mixture of vibrational ground and excited state atoms, parametrized by the probability $p=\langle n_1 \rangle$:
\begin{eqnarray}
\rho(0) &=& \left[(1-p)\,\ket{00}\bra{00} + \frac{p}{2}\, \ket{01}_s {}_s\bra{01} + \frac{p}{2}\,  \ket{01}_a {}_a\bra{01}\right]\nonumber\\
&\times& \ket{e\uparrow,g\downarrow}\bra{e\uparrow,g\downarrow},
\vspace*{4pt}
\end{eqnarray}
we numerically compute the fidelity of 40 repeated two-qubit SWAP gates. 
The fidelity of a single two-qubit gate degrades linearly with $p$. Moreover, upon performing repeated gates, the atoms are pumped in an incoherent, equally populated mixture of the symmetric and antisymmetric motional states, further deteriorating the gate fidelity (see Fig.~\ref{fig7}). This shows that ground-state cooling and/or entropy extraction are crucial tools for the architecture implementation.

\newpage
\section{Wavelengths for $^{173}$Yb and $^{87}$Sr}\label{app_E}
%
\begin{table}[h!]
\begin{center}
\begin{ruledtabular}
\begin{tabular}{lccccc}
\toprule
$\,$ & $\lambda \,({\rm nm})$ & $\alpha_e\, ({\rm a.u.})$ & $\alpha_g \,({\rm a.u.})$ & $\Gamma_g\, ({\rm s}^{-1})$  &  $\Gamma_e \, ({\rm s}^{-1})$ \\[2pt] \hline \\[-7pt] 
    Array & 465.4 & 382 & 382 & $53$  & $22$ \\ 
             & 759.4  & 186 & 186 & $7$ &  $11$ \\[5pt] 
    Transport\, &  570  & 0 & 257 & $29$ & $26$ \\
             &  980  & 0 & 164 & $6$ & $4$ \\[5pt] 
    OS & 555.8 & 49 &  
    375 \text{(vect.)} & $1300$ & $1$\\[5pt] 
    Clock & 578.4  & - & - & $4\times 10^{-2}$ & -   \\[5pt] 
    Detection  & 398.9 & $\approx\,$100  & -  & $1.8\times 10^8$ & $7\times10^{-6}$  \\[5pt]
    Quench  & 1388.8 & -  & -  & $ 2 $ & $ 1.7 \times 10^{6}$  \\
\bottomrule 
\end{tabular}
\end{ruledtabular}
\caption{
{\bf Summary of all required wavelengths for $^{173}$Yb in optical tweezers.} The AC polarizabilities $\alpha_{g,e}$ are reported in atomic units for $g$ and $e$ states \cite{Dzuba2009}. The scattering rates are given by considering the intensities required to achieve frequencies $\omega_{\rm tw}/2\pi=103\,{\rm kHz}$ and $\omega_{\rm tr}/2\pi=170\,{\rm kHz}$ with a beam waist of $1\,\mu\rm m$ for magic and transport tweezers, respectively. For the detection beam, $I/I_s=25$ is assumed. 
In the case of the OS beam, the total scattering rate $\Gamma_g$ is calculated considering a detuning $\delta_\textrm{OS}=45\,{\rm GHz}$ from the $^1S_0\rightarrow{^3P}_1\,F'=7/2$ resonance and $I/I_s = 5.3 \times 10^8$.}
\label{table_1}
\end{center}
\end{table}

\vspace{10pt}

\begin{table}[h!]
\begin{center}
\begin{ruledtabular}
\begin{tabular}{lccccc}
\toprule
$\,$ & $\lambda \,({\rm nm})$ & $\alpha_e\, ({\rm a.u.})$ & $\alpha_g \,({\rm a.u.})$ & $\Gamma_g\, ({\rm s}^{-1})$  &  $\Gamma_e \, ({\rm s}^{-1})$ \\[2pt] \hline \\[-7pt] 
    Array & 497.0  & 1350 & 1350 & $15.4$ &  $5.9$ \\
    & 813.4 & 284 & 284 & $3.4$  & $1.3$ \\[5pt] 
    Transport\, &  632.8  & 0 & 406 & $30$ & $15$ \\
             &  1666.6  & 0 & $\sim$ 200 & $2$ & $0.2$ \\[5pt] 
    OS & 689.3 & 1150 & 885 \text{(vect.)} & $40$ & $0.7$\\[5pt] 
    Clock & 698.4  & - & - & $6 \times 10^{-3}$ & -   \\[5pt] 
    Detection  & 460.7 & 1000  & -  & $2\times 10^8$ & $1\times10^{-4}$  \\
\bottomrule 
\end{tabular}
\end{ruledtabular}
\caption{
{\bf Summary of all required wavelengths for $^{87}$Sr in optical tweezers.} The AC polarizabilities $\alpha_{g,e}$ are reported in atomic units for $g$ and $e$ states \cite{Safronova2015}. The scattering rates are given considering the intensities required to achieve trap frequencies $\omega_{\rm tw}/2\pi=104\,{\rm kHz}$  and $\omega_{\rm tr}/2\pi=170\,{\rm kHz}$ with a beam waist of $1.3\,\mu\rm m$ for magic and transport tweezers respectively. For the detection beam, $I/I_s=25$ is assumed.  
In the case of the OS beam, the total scattering rate $\Gamma_g$ is calculated considering a detuning $\delta_\textrm{OS}=5\,{\rm GHz}$ from the $^1S_0\rightarrow{^3P}_1\,F'=11/2$ resonance and $I/I_s = 2.65 \times 10^9$.}
\label{table_2}
\end{center}
\end{table}
\clearpage
\bibliographystyle{ChemEurJ}
\bibliography{Tweezers_bib}

\providecommand{\url}[1]{\texttt{#1}}
\providecommand{\urlprefix}{}
\providecommand{\foreignlanguage}[2]{#2}
\providecommand{\Capitalize}[1]{\uppercase{#1}}
\providecommand{\capitalize}[1]{\expandafter\Capitalize#1}
\providecommand{\bibliographycite}[1]{\cite{#1}}
\providecommand{\bbland}{and}
\providecommand{\bblchap}{chap.}
\providecommand{\bblchapter}{chapter}
\providecommand{\bbletal}{et~al.}
\providecommand{\bbleditors}{editors}
\providecommand{\bbleds}{eds.}
\providecommand{\bbleditor}{editor}
\providecommand{\bbled}{ed.}
\providecommand{\bbledition}{edition}
\providecommand{\bbledn}{ed.}
\providecommand{\bbleidp}{page}
\providecommand{\bbleidpp}{pages}
\providecommand{\bblerratum}{erratum}
\providecommand{\bblin}{in}
\providecommand{\bblmthesis}{Master's thesis}
\providecommand{\bblno}{no.}
\providecommand{\bblnumber}{number}
\providecommand{\bblof}{of}
\providecommand{\bblpage}{page}
\providecommand{\bblpages}{pages}
\providecommand{\bblp}{p}
\providecommand{\bblphdthesis}{Ph.D. thesis}
\providecommand{\bblpp}{pp}
\providecommand{\bbltechrep}{Tech. Rep.}
\providecommand{\bbltechreport}{Technical Report}
\providecommand{\bblvolume}{volume}
\providecommand{\bblvol}{Vol.}
\providecommand{\bbljan}{January}
\providecommand{\bblfeb}{February}
\providecommand{\bblmar}{March}
\providecommand{\bblapr}{April}
\providecommand{\bblmay}{May}
\providecommand{\bbljun}{June}
\providecommand{\bbljul}{July}
\providecommand{\bblaug}{August}
\providecommand{\bblsep}{September}
\providecommand{\bbloct}{October}
\providecommand{\bblnov}{November}
\providecommand{\bbldec}{December}
\providecommand{\bblfirst}{First}
\providecommand{\bblfirsto}{1st}
\providecommand{\bblsecond}{Second}
\providecommand{\bblsecondo}{2nd}
\providecommand{\bblthird}{Third}
\providecommand{\bblthirdo}{3rd}
\providecommand{\bblfourth}{Fourth}
\providecommand{\bblfourtho}{4th}
\providecommand{\bblfifth}{Fifth}
\providecommand{\bblfiftho}{5th}
\providecommand{\bblst}{st}
\providecommand{\bblnd}{nd}
\providecommand{\bblrd}{rd}
\providecommand{\bblth}{th}
\begin{thebibliography}{10}

\bibitem{Jaksch1999}
D.~Jaksch, H.-J. Briegel, J.~I. Cirac, C.~W. Gardiner, P.~Zoller, \emph{Phys.
  Rev. Lett.} \textbf{1999}, \emph{82}, 1975--1978.

\bibitem{Brennen1999}
G.~K. Brennen, C.~M. Caves, P.~S. Jessen, I.~H. Deutsch, \emph{Phys. Rev.
  Lett.} \textbf{1999}, \emph{82}, 1060--1063.

\bibitem{Jaksch2000}
D.~Jaksch, J.~I. Cirac, P.~Zoller, S.~L. Rolston, R.~C\^ot\'e, M.~D. Lukin,
  \emph{Phys. Rev. Lett.} \textbf{2000}, \emph{85}, 2208--2211.

\bibitem{Briegel2000}
H.-J. Briegel, T.~Calarco, D.~Jaksch, J.~I. Cirac, P.~Zoller, \emph{J. Mod.
  Opt.} \textbf{2000}, \emph{47}, 415--451.

\bibitem{Calarco2000}
T.~Calarco, E.~Hinds, D.~Jaksch, J.~Schmiedmayer, J.~Cirac, P.~Zoller,
  \emph{Phys. Rev. A} \textbf{2000}, \emph{61}, 1--11.

\bibitem{Lukin2001}
M.~D. Lukin, M.~Fleischhauer, R.~Cote, L.~M. Duan, D.~Jaksch, J.~I. Cirac,
  P.~Zoller, \emph{Phys. Rev. Lett.} \textbf{2001}, \emph{87}, 037901.

\bibitem{Calarco2004}
T.~Calarco, U.~Dorner, P.~S. Julienne, C.~J. Williams, P.~Zoller, \emph{Phys.
  Rev. A} \textbf{2004}, \emph{70}, 012306.

\bibitem{Hayes2007}
D.~Hayes, P.~Julienne, I.~Deutsch, \emph{Phys. Rev. Lett.} \textbf{2007},
  \emph{98}, 1--4.

\bibitem{Saffman2010}
M.~Saffman, T.~G. Walker, K.~M\o{}lmer, \emph{Rev. Mod. Phys.} \textbf{2010},
  \emph{82}, 2313--2363.

\bibitem{Xia2015}
T.~Xia, M.~Lichtman, K.~Maller, A.~W. Carr, M.~J. Piotrowicz, L.~Isenhower,
  M.~Saffman, \emph{Phys. Rev. Lett.} \textbf{2015}, \emph{114}, 100503.

\bibitem{Wang2016}
Y.~Wang, A.~Kumar, T.-Y. Wu, D.~S. Weiss, \emph{Science} \textbf{2016},
  \emph{352}, 1562--1565.

\bibitem{Maller2015}
K.~M. Maller, M.~T. Lichtman, T.~Xia, Y.~Sun, M.~J. Piotrowicz, A.~W. Carr,
  L.~Isenhower, M.~Saffman, \emph{Phys. Rev. A} \textbf{2015}, \emph{92},
  022336.

\bibitem{Levine2018}
H.~{Levine}, A.~{Keesling}, A.~{Omran}, H.~{Bernien}, S.~{Schwartz}, A.~S.
  {Zibrov}, M.~{Endres}, M.~{Greiner}, V.~{Vuleti{\'c}}, M.~D. {Lukin},
  \emph{arXiv:1806.04682} \textbf{2018}.

\bibitem{DiVincenzo2000}
D.~P. DiVincenzo, \emph{Fortschr. Phys.} \textbf{2000}, \emph{48}, 771--783.

\bibitem{Mandel2003}
O.~Mandel, M.~Greiner, A.~Widera, T.~Rom, T.~W. H{\"a}nsch, I.~Bloch,
  \emph{Nature} \textbf{2003}, \emph{425}, 937.

\bibitem{Ratcliffe2018}
A.~K. Ratcliffe, R.~L. Taylor, J.~J. Hope, A.~R. Carvalho, \emph{Phys. Rev.
  Lett.} \textbf{2018}, \emph{120}, 220501.

\bibitem{Weitenberg2011a}
C.~Weitenberg, S.~Kuhr, K.~M\o{}lmer, J.~F. Sherson, \emph{Phys. Rev. A}
  \textbf{2011}, \emph{84}, 032322.

\bibitem{Anderlini2007}
M.~Anderlini, P.~J. Lee, B.~L. Brown, J.~Sebby-Strabley, W.~D. Phillips, J.~V.
  Porto, \emph{Nature} \textbf{2007}, \emph{448}, 452--456.

\bibitem{Kaufman2014}
A.~M. Kaufman, B.~J. Lester, C.~M. Reynolds, M.~L. Wall, M.~Foss-Feig, K.~R.~A.
  Hazzard, A.~M. Rey, C.~A. Regal, \emph{Science} \textbf{2014}, \emph{345},
  306--309.

\bibitem{Kaufman2015}
A.~M. Kaufman, B.~J. Lester, M.~Foss-Feig, M.~L. Wall, A.~M. Rey, C.~A. Regal,
  \emph{Nature} \textbf{2015}, \emph{527}, 208--211.

\bibitem{Gorshkov2010}
A.~V. Gorshkov, M.~Hermele, V.~Gurarie, C.~Xu, P.~S. Julienne, J.~Ye,
  P.~Zoller, E.~Demler, M.~D. Lukin, A.~M. Rey, \emph{Nature Phys.}
  \textbf{2010}, \emph{6}, 289--295.

\bibitem{Gerbier2010}
F.~Gerbier, J.~Dalibard, \emph{New J. Phys.} \textbf{2010}, \emph{12}, 033007.

\bibitem{Cazalilla2014}
M.~A. Cazalilla, A.~M. Rey, \emph{Rep. Prog. Phys.} \textbf{2014}, \emph{77},
  124401.

\bibitem{Ludlow2015}
A.~D. Ludlow, M.~M. Boyd, J.~Ye, E.~Peik, P.~O. Schmidt, \emph{Rev. Mod. Phys.}
  \textbf{2015}, \emph{87}, 637.

\bibitem{Daley2008}
A.~Daley, M.~Boyd, J.~Ye, P.~Zoller, \emph{Phys. Rev. Lett.} \textbf{2008},
  \emph{101}, 170504.

\bibitem{Gorshkov2009}
A.~Gorshkov, A.~Rey, A.~Daley, M.~Boyd, J.~Ye, P.~Zoller, M.~Lukin, \emph{Phys.
  Rev. Lett.} \textbf{2009}, \emph{102}, 1--4.

\bibitem{Shibata2009}
K.~Shibata, S.~Kato, A.~Yamaguchi, S.~Uetake, Y.~Takahashi, \emph{Applied
  Physics B} \textbf{2009}, \emph{97}, 753.

\bibitem{Daley2011b}
A.~Daley, \emph{Quantum Inf. Process.} \textbf{2011}, 865--884.

\bibitem{Scazza2014}
F.~Scazza, C.~Hofrichter, M.~H\"{o}fer, P.~C. {De Groot}, I.~Bloch,
  S.~F\"{o}lling, \emph{Nature Phys.} \textbf{2014}, \emph{10}, 779--784.

\bibitem{Zhang2014a}
X.~Zhang, M.~Bishof, S.~L. Bromley, C.~V. Kraus, M.~S. Safronova, P.~Zoller,
  A.~M. Rey, J.~Ye, \emph{Science} \textbf{2014}, \emph{345}, 1467.

\bibitem{Cappellini2014a}
G.~Cappellini, M.~Mancini, G.~Pagano, P.~Lombardi, L.~Livi, M.~{Siciliani de
  Cumis}, P.~Cancio, M.~Pizzocaro, D.~Calonico, F.~Levi, C.~Sias, J.~Catani,
  M.~Inguscio, L.~Fallani, \emph{Phys. Rev. Lett.} \textbf{2014}, \emph{113},
  120402.

\bibitem{Goban2018}
A.~Goban, R.~Hutson, G.~Marti, S.~Campbell, M.~Perlin, P.~Julienne, J.~D'Incao,
  A.~Rey, J.~Ye, \emph{arXiv:1803.11282} \textbf{2018}.

\bibitem{Pagano2014}
G.~Pagano, M.~Mancini, G.~Cappellini, P.~Lombardi, F.~Sch\"{a}fer, H.~Hu, X.-J.
  Liu, J.~Catani, C.~Sias, M.~Inguscio, L.~Fallani, \emph{Nature Phys.}
  \textbf{2014}, \emph{10}, 198--201.

\bibitem{Hinkley2013}
N.~Hinkley, J.~A. Sherman, N.~B. Phillips, M.~Schioppo, N.~D. Lemke,
  M.~Pizzocaro, C.~W. Oates, A.~D. Ludlow, \emph{Science} \textbf{2013},
  \emph{341}, 1215--1218.

\bibitem{Bloom2014}
B.~J. Bloom, T.~L. Nicholson, J.~R. Williams, S.~L. Campbell, M.~Bishof,
  X.~Zhang, W.~Zhang, S.~L. Bromley, J.~Ye, \emph{Nature} \textbf{2014},
  \emph{506}, 71--75.

\bibitem{Campbell2017}
S.~L. Campbell, R.~B. Hutson, G.~E. Marti, A.~Goban, N.~Darkwah~Oppong, R.~L.
  McNally, L.~Sonderhouse, J.~M. Robinson, W.~Zhang, B.~J. Bloom, J.~Ye,
  \emph{Science} \textbf{2017}, \emph{358}, 90--94.

\bibitem{Mancini2015}
M.~Mancini, G.~Pagano, G.~Cappellini, L.~Livi, M.~Rider, J.~Catani, C.~Sias,
  P.~Zoller, M.~Inguscio, M.~Dalmonte, L.~Fallani, \emph{Science}
  \textbf{2015}, \emph{349}, 1510--1513.

\bibitem{Hofrichter2016}
C.~Hofrichter, L.~Riegger, F.~Scazza, M.~H\"ofer, D.~R. Fernandes, I.~Bloch,
  S.~F\"olling, \emph{Phys. Rev. X} \textbf{2016}, \emph{6}, 021030.

\bibitem{Livi2017}
L.~F. Livi, G.~Cappellini, M.~Diem, L.~Franchi, C.~Clivati, M.~Frittelli,
  F.~Levi, D.~Calonico, J.~Catani, M.~Inguscio, L.~Fallani, \emph{Phys. Rev.
  Lett.} \textbf{2016}, \emph{117}, 220401.

\bibitem{Riegger2018}
L.~Riegger, N.~Darkwah~Oppong, M.~H\"ofer, D.~R. Fernandes, I.~Bloch,
  S.~F\"olling, \emph{Phys. Rev. Lett.} \textbf{2018}, \emph{120}, 143601.

\bibitem{Endres2016}
M.~Endres, H.~Bernien, A.~Keesling, H.~Levine, E.~R. Anschuetz, A.~Krajenbrink,
  C.~Senko, V.~Vuletic, M.~Greiner, M.~D. Lukin, \emph{Science} \textbf{2016},
  \emph{354}, 1024.

\bibitem{Labuhn2016}
H.~Labuhn, D.~Barredo, S.~Ravets, S.~de~L{\'e}s{\'e}leuc, T.~Macr{\`\i},
  T.~Lahaye, A.~Browaeys, \emph{Nature} \textbf{2016}, \emph{534}, 667.

\bibitem{Barredo2016}
D.~Barredo, S.~De~L{\'e}s{\'e}leuc, V.~Lienhard, T.~Lahaye, A.~Browaeys,
  \emph{Science} \textbf{2016}, \emph{354}, 1021.

\bibitem{Bernien2017}
H.~Bernien, S.~Schwartz, A.~Keesling, H.~Levine, A.~Omran, H.~Pichler, S.~Choi,
  A.~S. Zibrov, M.~Endres, M.~Greiner, V.~Vuleti{\'c}, M.~D. Lukin,
  \emph{Nature} \textbf{2017}, \emph{551}, 579.

\bibitem{Liu2018}
L.~R. Liu, J.~D. Hood, Y.~Yu, J.~T. Zhang, N.~R. Hutzler, T.~Rosenband, K.-K.
  Ni, \emph{Science} \textbf{2018}, \emph{360}, 900--903.

\bibitem{Ni2018}
K.-K. Ni, T.~Rosenband, D.~D. Grimes, \emph{Chem. Sci.} \textbf{2018},
  \emph{9}, 6830--6838.

\bibitem{Olmschenk2007}
S.~Olmschenk, K.~C. Younge, D.~L. Moehring, D.~N. Matsukevich, P.~Maunz,
  C.~Monroe, \emph{Phys. Rev. A} \textbf{2007}, \emph{76}.

\bibitem{Daley2011a}
A.~J. Daley, J.~Ye, P.~Zoller, \emph{Eur. Phys. J. D} \textbf{2011}, \emph{65},
  207--217.

\bibitem{Porsev2004}
S.~Porsev, A.~Derevianko, E.~Fortson, \emph{Phys. Rev. A} \textbf{2004},
  \emph{69}, 021403.

\bibitem{ScazzaPhD}
F.~Scazza, \bblphdthesis{}, Ludwig-Maximilians-Universit{\"a}t M{\"u}nchen,
  \textbf{2015}.

\bibitem{Boyd2007}
M.~Boyd, T.~Zelevinsky, A.~Ludlow, S.~Blatt, T.~Zanon-Willette, S.~Foreman,
  J.~Ye, \emph{Phys. Rev. A} \textbf{2007}, \emph{76}, 022510.

\bibitem{Pagano2015a}
G.~Pagano, \bblphdthesis{}, Scuola Normale Superiore di Pisa, \textbf{2015}.

\bibitem{Reichenbach2007}
I.~Reichenbach, I.~H. Deutsch, \emph{Phys. Rev. Lett.} \textbf{2007},
  \emph{99}, 123001.

\bibitem{Ozeri2007}
R.~Ozeri, W.~M. Itano, R.~B. Blakestad, J.~Britton, J.~Chiaverini, J.~D. Jost,
  C.~Langer, D.~Leibfried, R.~Reichle, S.~Seidelin, J.~H. Wesenberg, D.~J.
  Wineland, \emph{Phys. Rev. A} \textbf{2007}, \emph{75}.

\bibitem{Uys2010}
H.~Uys, M.~J. Biercuk, A.~P. VanDevender, C.~Ospelkaus, D.~Meiser, R.~Ozeri,
  J.~J. Bollinger, \emph{Phys. Rev. Lett.} \textbf{2010}, \emph{105}.

\bibitem{Busch1998}
T.~Busch, B.~G. Englert, K.~Rzazewski, M.~Wilkens, \emph{Found. Phys.}
  \textbf{1998}, \emph{28}, 549--559.

\bibitem{Hofer2015}
M.~H\"ofer, L.~Riegger, F.~Scazza, C.~Hofrichter, D.~R. Fernandes, M.~M.
  Parish, J.~Levinsen, I.~Bloch, S.~F\"olling, \emph{Phys. Rev. Lett.}
  \textbf{2015}, \emph{115}, 265302.

\bibitem{Yamamoto2016}
R.~Yamamoto, J.~Kobayashi, T.~Kuno, K.~Kato, Y.~Takahashi, \emph{New J. Phys.}
  \textbf{2016}, \emph{18}, 023016.

\bibitem{Nemitz2016}
N.~Nemitz, T.~Ohkubo, M.~Takamoto, I.~Ushijima, M.~Das, N.~Ohmae, H.~Katori,
  \emph{Nat. Photonics} \textbf{2016}, \emph{10}, 258--261.

\bibitem{Brown2017}
R.~C. Brown, N.~B. Phillips, K.~Beloy, W.~F. McGrew, M.~Schioppo, R.~J. Fasano,
  G.~Milani, X.~Zhang, N.~Hinkley, H.~Leopardi, T.~H. Yoon, D.~Nicolodi, T.~M.
  Fortier, A.~D. Ludlow, \emph{Phys. Rev. Lett.} \textbf{2017}, \emph{119},
  253001.

\bibitem{Curtis2001}
E.~A. Curtis, C.~W. Oates, L.~Hollberg, \emph{Phys. Rev. A} \textbf{2001},
  \emph{64}, 031403.

\bibitem{Leibfried2003}
D.~Leibfried, R.~Blatt, C.~Monroe, D.~Wineland, \emph{Rev. Mod. Phys.}
  \textbf{2003}, \emph{75}, 281--324.

\bibitem{Miranda2014}
M.~Miranda, R.~Inoue, Y.~Okuyama, A.~Nakamoto, M.~Kozuma, \emph{Phys. Rev. A}
  \textbf{2015}, \emph{91}, 063414.

\bibitem{Fogarty2018}
T.~{Fogarty}, L.~{Ruks}, J.~{Li}, T.~{Busch}, \emph{arXiv: 1806.08506}
  \textbf{2018}.

\bibitem{Briegel2001}
H.~J. Briegel, R.~Raussendorf, \emph{Phys. Rev. Lett.} \textbf{2001},
  \emph{86}, 910.

\bibitem{Briegel2009}
H.~J. Briegel, D.~E. Browne, W.~D{\"u}r, R.~Raussendorf, M.~Van~den Nest,
  \emph{Nature Phys.} \textbf{2009}, \emph{5}, 19.

\bibitem{Raussendorf2003}
R.~Raussendorf, D.~E. Browne, H.~J. Briegel, \emph{Phys. Rev. A} \textbf{2003},
  \emph{68}, 022312.

\bibitem{Foss-Feig2010}
M.~Foss-Feig, M.~Hermele, A.~M. Rey, \emph{Phys. Rev. A} \textbf{2010},
  \emph{81}, 051603.

\bibitem{Ono2018}
K.~Ono, J.~Kobayashi, Y.~Amano, K.~Sato, Y.~Takahashi, \emph{arXiv:1810.00536}
  \textbf{2018}.

\bibitem{Cooper2018}
A.~Cooper, J.~P. Covey, I.~S. Madjarov, S.~G. Porsev, M.~S. Safronova,
  M.~Endres, \emph{arXiv:1810.06537} \textbf{2018}.

\bibitem{Norcia2018}
M.~Norcia, A.~Young, A.~Kaufman, \emph{arXiv:1810.06626} \textbf{2018}.

\bibitem{Saskin2018}
S.~Saskin, J.~Wilson, B.~Grinkemeyer, J.~Thompson, \emph{arXiv:1810.10517}
  \textbf{2018}.

\bibitem{Covey2018}
J.~P. Covey, I.~S. Madjarov, A.~Cooper, M.~Endres, \emph{arXiv:1811.06014}
  \textbf{2018}.

\bibitem{doi:10.1063/1.1664991}
H.~R. Lewis, W.~B. Riesenfeld, \emph{J. Math. Phys.} \textbf{1969}, \emph{10},
  1458--1473.

\bibitem{PhysRevA.54.5378}
K.~B. Moller, T.~G. Jorgensen, J.~P. Dahl, \emph{Phys. Rev. A} \textbf{1996},
  \emph{54}, 5378--5385.

\bibitem{Stellmer2011}
S.~Stellmer, R.~Grimm, F.~Schreck, \emph{Phys. Rev. A} \textbf{2011},
  \emph{84}, 043611.

\bibitem{Taie2010}
S.~Taie, Y.~Takasu, S.~Sugawa, R.~Yamazaki, T.~Tsujimoto, R.~Murakami,
  Y.~Takahashi, \emph{Phys. Rev. Lett.} \textbf{2010}, \emph{105}, 190401.

\bibitem{Dzuba2009}
V.~A. Dzuba, A.~Derevianko, \emph{J. Phys. B: At. Mol. Opt. Phys.}
  \textbf{2009}, \emph{43}, 11.

\bibitem{Safronova2015}
M.~S. Safronova, Z.~Zuhrianda, U.~I. Safronova, C.~W. Clark, \emph{Phys. Rev.
  A} \textbf{2015}, \emph{92}, 040501.

\end{thebibliography}




\end{document}